\documentclass[twocolumn,traditabstract]{aa}  

\usepackage{amsmath}
\usepackage{amssymb}
\usepackage{fixltx2e}
\usepackage[english]{babel}
\usepackage{graphicx}
\usepackage{epstopdf}
\usepackage{epsf,color}
\usepackage[mathscr]{eucal}
\usepackage{amsmath}
\usepackage{amssymb,amsfonts}
\usepackage{graphicx}
\usepackage{txfonts}
\usepackage{dsfont}
\definecolor{Mygreen}{rgb}{0.75, 0.0, 0.0}
\definecolor{Mypink}{rgb}{1.0, 0.0, 0.5}
\definecolor{Myred}{rgb}{0.7, 0.0, 0.0}
\usepackage[breaklinks, citecolor=blue, linkcolor=Myred, urlcolor=Myred, colorlinks=true]{hyperref}
\usepackage{float} 
\usepackage{color}
\usepackage[usenames,dvipsnames]{xcolor}

\usepackage{ulem}
\usepackage{hyperref}
\usepackage{mwe,tikz}
\usepackage[percent]{overpic}
\usepackage{booktabs}
\usepackage{cuted}

\usepackage{microtype}

\usepackage{natbib}
\bibpunct{(}{)}{;}{a}{}{,}

\begin{document}

\title{PITSZI: Probing Intra-cluster medium Turbulence with Sunyaev-Zel'dovich Imaging}
\subtitle{Application to the triple merging cluster MACS~J0717.5+3745}

\author{
R.~Adam \inst{\ref{OCA}}
\and  T.~Eynard-Machet\inst{\ref{OCA}}
\and  I.~Bartalucci\inst{\ref{Milano}}
\and  D.~Cherouvrier\inst{\ref{LPSC}}
\and  N.~Clerc\inst{\ref{IRAP}}
\and  L.~Di Mascolo\inst{\ref{OCA}}
\and  S.~Dupourqué\inst{\ref{IRAP}}
\and  C.~Ferrari\inst{\ref{OCA}}
\and  J.-F.~Macías-Pérez\inst{\ref{LPSC}}
\and  E.~Pointecouteau\inst{\ref{IRAP}}
\and  G.W.~Pratt\inst{\ref{CEA}}
}

\offprints{Rémi Adam (\url{remi.adam@oca.eu})}

\institute{
Universit\'e C\^ote d'Azur, Observatoire de la C\^ote d'Azur, CNRS, Laboratoire Lagrange, France
  \label{OCA}
  \and
INAF, IASF-Milano, Via A. Corti 12, 20133 Milano, Italy
  \label{Milano}
  \and
  Laboratoire de Physique Subatomique et de Cosmologie, Universit\'e Grenoble Alpes, CNRS/IN2P3, 53, avenue des Martyrs, Grenoble, France
  \label{LPSC}
    \and
IRAP, CNRS, Université de Toulouse, CNES, UT3-UPS, (Toulouse), France
  \label{IRAP}
  \and
Universit\'e Paris-Saclay, Universit\'e Paris Cit\'e, CEA, CNRS, AIM, F-91191, Gif-sur-Yvette, France (2022)
  \label{CEA}
}

\date{Received \today \ / Accepted --}
\abstract 
{Turbulent gas motions are expected to dominate the non-thermal energy budget of the intracluster medium (ICM). The measurement of pressure fluctuations from high angular resolution Sunyaev-Zel'dovich imaging opens a new avenue to study ICM turbulence, complementary to X-ray density fluctuation measures. We develop a methodological framework designed to optimally extract information on the ICM pressure fluctuation power spectrum statistics, and publicly release the associated software named \texttt{PITSZI} (Probing ICM Turbulence from Sunyaev-Zel'dovich Imaging). We apply this tool to the New IRAM KIDs Array (NIKA) data of the merging cluster MACS~J0717.5+3745 to measure its pressure fluctuation power spectrum at high significance, and to investigate the implications for its nonthermal content. Depending on the choice of the radial pressure model and the details of the applied methodology, we measure an energy injection scale $L_{\rm inj} \sim 800$ kpc. The power spectrum normalization corresponds to a characteristic amplitude reaching $A_{\delta P / \bar{P}}(k_{\rm peak}) \sim 0.4$. These results are are obtained assuming that the ICM of MACS~J0717.5+3745 can be described as pressure fluctuations on top of a single (smooth) halo, and are dominated by systematics due to the choice of the radial pressure model. Using simulations, we estimate that fitting a radial model to the data can suppress the observed fluctuations by up to $\sim 50$\%, while a poorly representative radial model can induce spurious fluctuations, which we also quantify. Assuming standard scaling relations between the pressure fluctuations and turbulence, we find that MACS~J0717.5+3745 presents a turbulent velocity dispersion $\sigma_v \sim 1200$ km/s, a kinetic to kinetic plus thermal pressure fraction $P_{\rm kin} / P_{\rm kin+th} \sim 20\%$, and we estimate the hydrostatic mass bias to $b_{\rm HSE} \sim 0.3-0.4$. Our results are in excellent agreement with alternative measurements obtained from X-ray surface brightness fluctuations, and in agreement with the fluctuations being adiabatic in nature.}

\titlerunning{\texttt{PITSZI}: Probing ICM Turbulence from Sunyaev-Zel'dovich Imaging}
\authorrunning{R. Adam et al.}
\maketitle
\setcounter{secnumdepth}{3}

\section{Introduction}\label{sec:introduction}
In the 1970s, X-ray observations revealed the existence of a diffuse hot gas component that dominates the baryon content of galaxy clusters, known as the intracluster medium \citep[ICM ; see][for a review]{Sarazin1986}. Since then, the ICM has been extensively studied using X-ray observations, but also in the millimeter via the Sunyaev-Zel'dovich effect \citep[SZ;][]{Sunyaev1970,Sunyaev1972}. Today, we know that the physical properties of the ICM are driven by the gravitational collapse of the surrounding cosmic web, in which the kinetic energy of the infalling gas is converted into heat primarily through shocks and turbulent cascades \citep{Kravtsov2012}. Consequently, the ICM is essentially thermal. However, a non-negligible fraction of the energy is also channeled into a non-thermal component in the form of turbulence, magnetic fields, and cosmic rays. While the thermal state of galaxy clusters is now well understood, the non-thermal content is not. In particular, turbulent gas motions are expected to dominate the nonthermal ICM energy budget \citep[e.g.,][]{Vazza2016} and are presumably key to explain cosmic-ray (re)acceleration and magnetic field amplification \citep{Brunetti2014,Donnert2018}. Therefore, the characterization of ICM turbulence is becoming crucial to acquire a comprehensive view of the physical processes involved in the assembly of massive halos and utilizing them for high-precision cluster cosmology.

Indeed, robust and precise total mass estimates are essential for the use of the galaxy cluster population as a cosmological probe \citep{Pratt2019}. As the dominant contribution to the nonthermal pressure support, turbulence is expected to play a leading contribution to the hydrostatic mass bias that affect cluster masses derived from the hydrostatic equilibrium assumption \citep{Angelinelli2020}. The hydrostatic mass bias has been widely examined in the literature through both observational data, especially via the calibration of mass scaling relations, and numerical simulations \citep[e.g.,][]{Hoekstra2015,Smith2016,Eckert2019,Gianfagna2021,Munoz2024}, in particular in light of the tension between cosmic microwave background-based and cluster-based cosmology \citep{PlanckXX2014}. Nevertheless, the understanding of its physical origin and its complete characterization is still an open issue. In fact, the hydrostatic mass bias is currently one of the major limitations for the use of ICM-based observations as a cosmological probe \citep[e.g.,][]{PlanckXXIV2016,Pratt2019}.

Turbulence is also a key ingredient to explain the diffuse radio emission that is widely observed in galaxy clusters today, but not yet fully understood. We can differentiate between radio relics, which are elongated structures located at the periphery of clusters, and radio halos, which coincide spatially with the thermal gas \citep{vanWeeren2019}. While relics are believed to trace merger shock acceleration \citep{vanWeeren2010}, turbulence is often invoked to explain the failure of the simple diffuse shock acceleration model \citep{Fujita2015}. Radio halos are generally further divided into mini-halos (in relaxed cool-core clusters), megaparsec-size giant radio halos (in disturbed objects), and mega-halos, which extend out to the extreme periphery. In all cases, observations suggest that turbulence is a key player to reaccelerate synchrotron emitting relativistic electrons \citep{Brunetti2014}. In this scenario, mini-halos may be powered by AGN feedback or sloshing motions \citep{Mazzotta2008}, giant-halos by the energy dissipated in mergers \citep{Cassano2010}, and mega-halos by large scale structure formation \citep{Beduzzi2023}. However, the details of the mechanisms involved are still poorly understood and the origin of the particles re-accelerated by turbulence is debated \citep[e.g.,][]{Pinzke2017,Adam2021}. Characterizing the connection between these nonthermal processes and turbulence, in different environments, is therefore a key issue to develop a thorough characterization of the plasma physics at play.

Ideally, turbulence should be observed by directly probing the ICM velocity field \citep{Simionescu2019}. This will become routine with high resolution X-ray spectroscopy experiments \citep{XRISM2020,Barret2020}, following on from the early results from the \textit{Hitomi} satellite towards the core of the Perseus cluster \citep{Hitomi2016}. Alternatively, turbulence can be probed indirectly through fluctuations of ICM thermodynamic quantities, assuming these are being produced by the turbulent velocity field. This approach relies on the calibration of the power spectrum of the target ICM fluctuation with the turbulent Mach number, as studied in numerical simulations \citep{Zhuravleva2014,Gaspari2014,Mohapatra2020,Mohapatra2021,Mohapatra2022,Simonte2022,Zhuravleva2023}. To date, such studies have been conducted through the statistics of X-ray surface brightness and pressure fluctuations obtained relative to a smooth model \citep[e.g.,][]{Schuecker2004,Churazov2012,Dupourque2023,Dupourque2024,Heinrich2024}. However, the physical interpretation of these fluctuations as being generated by turbulence is not straightforward from X-rays only. They may also arise from structures that are independent of turbulence \citep[e.g., clumping, sloshing features and shocks; see][for discussions]{Churazov2012,Dupourque2023}. Moreover, the thermodynamic nature of the fluctuations may depend on the ICM environment\footnote{Pressure and density perturbations are related via $\delta P/P = \Gamma \delta \rho/\rho$, with $\Gamma = [0, 1, 5/3]$ for isobaric, isothermal, and adiabatic fluctuations.}. For instance, isobaric fluctuations may arise from slow sloshing motions or gas bubbles induced by AGN feedback, while adiabatic fluctuations are associated with more vigorous motions such as with weak shocks or sonic turbulence \citep[see, e.g.,][]{Zhuravleva2018}. These regimes cannot be accessed directly through X-ray surface brightness fluctuations alone, since these probe only density perturbations.

The SZ effect offers an independent and complementary probe of the ICM \citep[see][for reviews]{Birkinshaw1999,Mroczkowski2019}. Unlike the X-ray emission, which traces gas density and temperature (with imaging and spectroscopy, respectively), the SZ effect directly measures the thermal pressure\footnote{It may also measure the ICM velocity from the kinetic SZ effect (kSZ) in individual systems, but this remains extremely challenging \citep{Sayers2013,Adam2017a}.}. Furthermore, the SZ surface brightness is independent of the redshift, which makes it very attractive for the characterization of distant objects, provided that sufficiently high angular resolution and sensitivity are available. As such, the SZ surface brightness fluctuations are obvious tracers to investigate the impact of turbulence on the ICM physics. To date, only three such studies have been conducted on individual objects \citep{Khatri2016,Romero2023,Romero2024a}, albeit with a limited signal-to-noise-ratio. \citet{Romero2024b} also undertook a forecast study on constraints from SZ surface brightness fluctuations in galaxy clusters. 

Here, we aim to develop a methodological framework that efficiently extracts information on the ICM pressure fluctuation power spectrum from SZ observations. We also release the corresponding software, named \texttt{PITSZI} (Probing ICM Turbulence from Sunyaev-Zel'dovich Imaging), to the public. We apply this tool to NIKA observations of the triple merging cluster MACS~J0717.5+3745, to measure its pressure fluctuation power spectrum and explore the implications for its nonthermal energy content.

The present article is structured as follows. Section~\ref{sec:overview} gives an overview of the \texttt{PITSZI} software. In Section~\ref{sec:data}, we present the NIKA data of the cluster MACS~J0717.5+3745, which we use as a test case throughout the paper. The modeling of the pressure distribution and its SZ observable, accounting for instrumental effects and noise, is discussed in Section~\ref{sec:modeling}. Section~\ref{sec:inference} presents the methodology used to infer constraints on the pressure profile and the pressure fluctuation power spectrum. We discuss the results highlighting the implications for the nonthermal ICM physics of MACS~J0717.5+3745 in Section~\ref{sec:physics}. Section~\ref{sec:Summary_and_conclusions} gives a summary of the paper and details our conclusions. Several appendices complement the paper by quantifying the relevant sources of uncertainty and systematic effects, and discuss how background and foreground contamination may be accounted for. Throughout the paper, we assume a flat $\Lambda$CDM cosmology with $H_0 = 70$ km s$^{-1}$ Mpc$^{-1}$ and $\Omega_{\rm m} = 0.3$. At the redshift of MACS~J0717.5+3745 ($z = 0.546$), 1 arcmin corresponds to 395 kpc.

\section{General overview and structure of the \texttt{PITSZI} software}\label{sec:overview}
\texttt{PITSZI} is a \texttt{python}-based code\footnote{The code, together with several example notebooks, is available at \url{https://github.com/remi-adam/pitszi}.} designed to model the ICM pressure distribution in galaxy clusters. It allows us to generate SZ image mocks and derive constraints on the power spectrum of the pressure fluctuations and its implications to the nonthermal ICM physics. Before entering into the details of the modeling and analysis framework, this section gives a brief overview of the code.

\begin{figure*}
        \centering
        \includegraphics[width=0.99\textwidth]{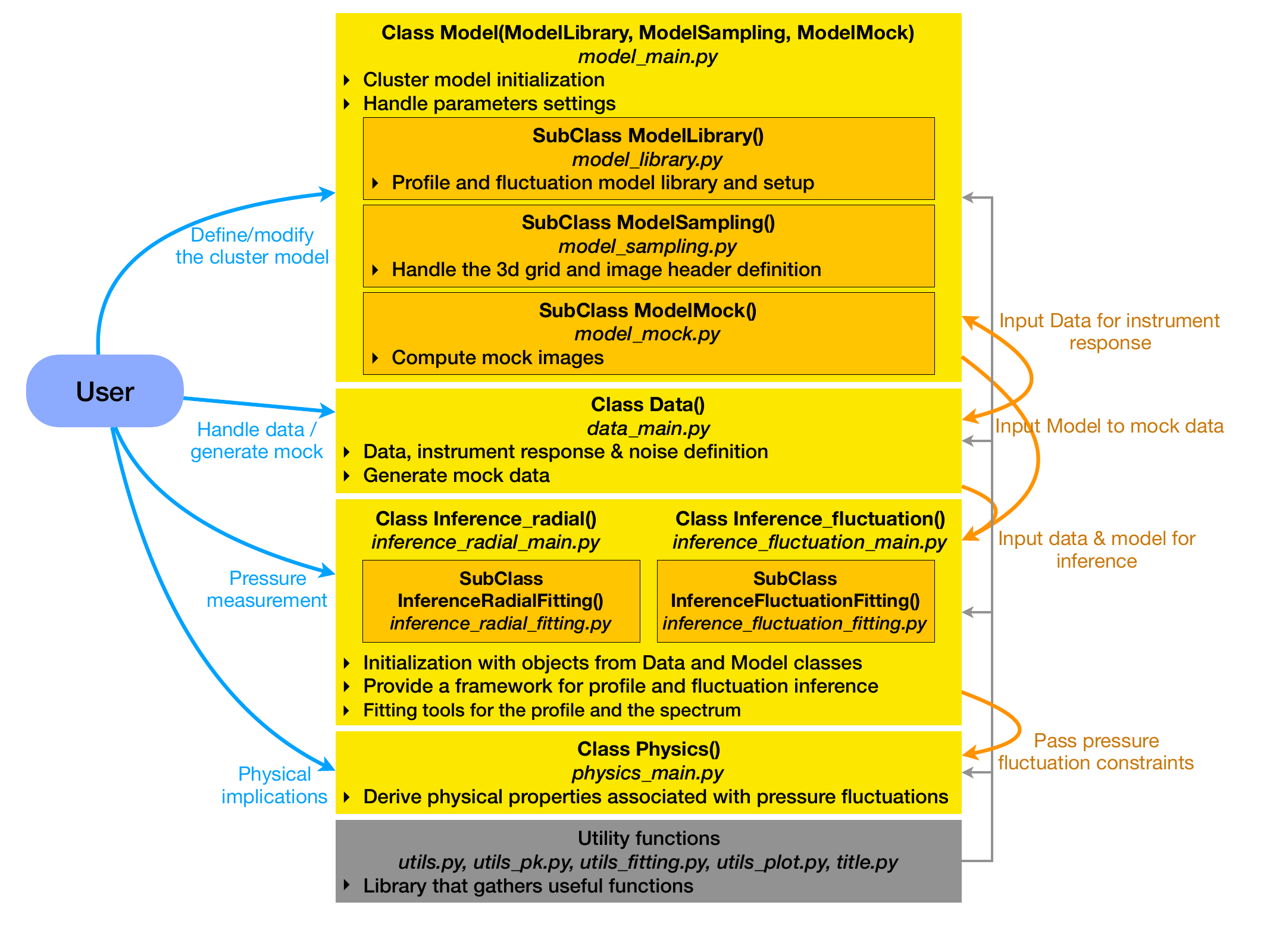}
        \caption{Overview of the code structure, user interfaces, and links between the different components. The yellow boxes display the main classes and the inner orange boxes indicate the dependence on the main sub-classes, with class names indicated at the top together with the name of the main file where this is defined. The grey box gives the name of the utility function files. The blue arrows indicate the interfaces with the user. The orange arrows indicate how the different modules are inputs to one another. The grey arrows show how utility functions are used by the classes.}
\label{fig:overview}
\end{figure*}

\subsection{Overview and key ingredients}
Modeling of the pressure distribution requires a  description of the ICM pressure profile, to which pressure fluctuations are added via the modeling of their power spectrum. Given a three dimensional grid sampling along the sky coordinates and the line-of-sight, the pressure is integrated to produce a mock SZ image. The SZ image may then be convolved with an instrument response function (beam smoothing and data reduction transfer function) and realistic noise may be added to mimic the data obtained from a given telescope. \texttt{PITSZI} was developed in the context of the analysis of NIKA and NIKA2 data \citep{Adam2014,Catalano2014,Adam2018a}, but it can in principle be applied to data obtained from any SZ imaging instrument (e.g., Planck, ACT, SPT, Bolocam, MUSTANG, MUSTANG2). 

To extract constraints on the pressure fluctuation power spectrum in three dimensions, one must first compute the power spectrum of the SZ image in two dimensions after accounting for the radial pressure model and given a region of interest (e.g., within $R_{500}$). This power spectrum is then compared to a model, which can be obtained via different methods, accounting for projection effects, the instrument response function, the masking of bad pixels, the noise, and possible astrophysical contaminants. To avoid noise bias and mitigate systematic effects in the data, one may use the cross power spectra of two independent data (sub)sets, if available.

With the constraints on the pressure fluctuation power spectrum in hand, \texttt{PITSZI} implements several approaches from the literature to derive the nonthermal properties of the ICM, such as the turbulent velocity or the nonthermal to thermal energy ratio.

\subsection{Code structure}
The structure of the code is illustrated in Figure~\ref{fig:overview}. \texttt{PITSZI} consists of four main classes.
\begin{enumerate}
\item The class \textit{Model} is designed to describe the cluster object and the pressure sampling. In addition to basic properties such as sky coordinates, mass and redshift, it includes the description of the pressure profile and the pressure fluctuations as the key physical properties.
\item The class \textit{Data} can be used to construct a data object, defined according to an SZ image, its astrometry information and the instrument response function and noise properties. This class also implements functions to handle the noise properties and generate noise Monte Carlo realizations.
\item The classes \textit{InferenceRadial} and \textit{InferenceFluctuation} are dedicated to extraction of constraints on the radial and fluctuation model parameters. Several methods to compute the SZ fluctuation power spectrum from a given pressure model are implemented, in addition to two different fitting methods based on non-linear least squares or Markov Chain Monte Carlo.
\item The class \textit{Physics} derives physical constraints on the nonthermal ICM physics given the pressure fluctuation power spectrum. It implements various results from the literature that connect turbulence to pressure fluctuations.
\end{enumerate}
In addition, libraries of utility functions, including those related to power spectrum analysis, are used by the different classes.

\subsection{Sampling}
The ICM pressure is sampled in real space by defining a box centered at the cluster redshift and on a reference map coordinate. This setup assumes that the cluster extent is sufficiently small on the sky to approximate the box as a simple pixelated rectangular parallelepiped. The line of sight is gridded given the pixel resolution and the box size along the line-of-sight (typically $3 \times R_{500}$). The projected grid may be provided via a header (e.g. corresponding to real data) or by defining the pixel resolution and the size of the field of view along longitude and latitude directions.

Fast Fourier transform definitions of the spatial frequency space, $\vec{k}$, follow that of the \texttt{numpy.fft} package\footnote{\url{https://numpy.org/doc/stable/reference/routines.fft.html}}, which is used in \texttt{PITSZI}. Note that the $k$-space does not include modes for which $\left|\vec{k} \right| > {\rm min}\left({\rm max}(k_1), {\rm max}(k_2), {\rm max}(k_3)\right)$, where $k_1, k_2, k_3$ are the spatial frequency along the plane of the sky and line-of-sight, respectively, so that the fluctuations would not be isotropic at these scales \citep[see][and in particular their figure 1]{Ponthieu2011}, but they are usually well below the limit given by the telescope angular resolution so that this is not relevant in practice.

\section{Data: MACS~J0717.5+3745 as observed with NIKA}\label{sec:data}
In this paper, we apply the \texttt{PITSZI} code to the cluster MACS~J0717.5+3745, as imaged by the NIKA camera. This section discusses the target choice, presents the NIKA data and reviews the possible contaminants that are accounted for in the analysis.

\subsection{MACS~J0717.5+3745}
MACS~J0717.5+3745 is one of the most complex and dynamically active galaxy clusters known. It is a triple merging system at redshift $z = 0.546$, involving at least four subclusters \citep{Ma2009}, and one of the most massive clusters in the Universe \citep[$M(< 1 \ {\rm Mpc}) \sim 2 \times 10^{15} $M$_{\odot}$,][]{Jauzac2018}. MACS~J0717.5+3745 is characterized by the presence of very large amounts of both thermal and nonthermal energy. Its ICM was found to be extremely hot, reaching up to $25$ keV \citep{Adam2017b}, and it hosts one of the most powerful known radio halos \citep{Rajpurohit2021a} and a complex radio relic \citep{Rajpurohit2021b}. Recently, the analysis of ICM density fluctuations with \textit{Chandra} data in a sample of 80 clusters found that  MACS~J0717.5+3745 displayed the highest characteristic velocities in the sample \citep{Heinrich2024}. In addition, MACS~J0717.5+3745 is currently the only individual cluster in which a kinetic SZ signal has been detected \citep{Mroczkowski2012,Sayers2013,Adam2017a}. As a well-studied system, the SZ images of MACS~J0717.5+3745 are among the best available, both in terms of signal-to-noise ratio and angular resolution. Given these properties, this cluster is an ideal test and benchmark target to search for pressure fluctuations via SZ mapping. However, as will be further discussed in Section~\ref{sec:physics}, we note that the interpretation of the results in terms of turbulence may not be straightforward given the complexity of the system. Nevertheless, it remains an excellent target for addressing the methodology that we develop here, which is the main goal of the paper.

\subsection{Data}
We use the 150 GHz NIKA data obtained from the IRAM 30m telescope under projects 237-13 and 222-14 (displayed in Figure~\ref{fig:example_images_data}). These correspond to about 13h of unflagged data taken under good atmospheric conditions. Calibration uncertainties were estimated to be 7\% at 150 GHz and the Gaussian beam FWHM (full width at half maximum) was measured to be 18.2 arcsec. The data processing induced transfer function, that filters scales larger than the field-of-view ($\sim 2$ arcmin) was estimated following \cite{Adam2015}. Noise Monte Carlo realizations were generated following the method described in \cite{Adam2016}. These data have been made publicly available in \cite{Adam2018b}\footnote{\url{https://lpsc.in2p3.fr/NIKA2LPSZ/nika2sz.release.php}}. We refer to \cite{Adam2017a,Adam2017b} for further details regarding the data reduction.

\subsection{Contaminants}
In addition to the thermal SZ signal and the target pressure fluctuations, a few astrophysical contaminants may affect the data. We account for these contaminants as follows. The atmospheric and instrumental noise will be discussed separately in Section~\ref{sec:modeling}.

First of all, MACS~J0717.5+3745 is known to be contaminated by kinetic SZ signal. We use the best fit template obtained in \cite{Adam2017a} as a correction. However, following their recommendation, since the model is degenerate and presents large uncertainties, we only used it to test the impact of the kinetic SZ signal as a systematic effect. We refer to Appendix~\ref{app:kSZ_systematics} for details regarding  the quantification of this effect.

The NIKA 150 GHz images are affected by resolved individual radio and infrared sources. We used the point source model built in \cite{Adam2017a} to correct for these. Additionally, we built a conservative point source mask by thresholding the point source template at 0.1 mJy/beam. We will apply this mask to the data to test the systematic effects due to possible remaining point source residuals.

In addition to the sources directly detected in the data, we expect the SZ signal to be affected by a contribution from unresolved galaxies below the detection threshold, that constitute the cosmic infrared background (CIB). We followed the methodology developed by \cite{Bethermin2017} to generate 1000 mock CIB realizations which were used to account for this contaminant in the following. We refer to Appendix~\ref{app:unresolved_background_cib} for further details regarding the construction of these mocks.

The contribution from undetected lower mass halos in the field of view may also potentially affect the data. However, we verify in Appendix~\ref{app:unresolved_background_halos}, using simulations, that this signal is negligible in the present case.

Finally, the diffuse galactic emission and the primary CMB anisotropies may also affect the data and should be accounted for if relevant. In the case of the NIKA data, however, both of these contaminants were shown to be negligible given the image sensitivity and the angular scales involved \citep{Adam2016,Adam2017a}.

\section{Modeling the pressure fluctuations and its Sunyaev-Zel'dovich observable}\label{sec:modeling}

This Section discusses how the pressure distribution is modeled in \texttt{PITSZI}, how it is used to generate mock SZ images, and how the observables are related to the underlying pressure fluctuation power spectrum.

\subsection{Pressure distribution and projection}\label{sec:Pressure_distribution_and_projection}
Let us first consider the modeling of the pressure and its projection to obtain the requested SZ observable and the power spectrum of its fluctuations.

\subsubsection{The Sunyaev-Zel'dovich effect}
The thermal SZ effect is due to the inverse Compton scattering of CMB photons by the hot ICM \citep{Sunyaev1970,Sunyaev1972}. The SZ surface brightness $\Delta I_{\nu}$, at frequency $\nu$, relative to the CMB specific intensity $I_0$, can be expressed as \citep[see][for a review]{Mroczkowski2019}
\begin{equation}
\frac{\Delta I_{\nu}}{I_0} = f(\nu, T) \times y,
\end{equation}
where $f(\nu, T)$ is the characteristic SZ spectrum and $y$ the Compton parameter, which gives the amplitude of the SZ effect. The SZ spectrum slightly depends on the temperature $T$ for very hot gas in the relativistic regime \citep{Pointecouteau1998,Itoh1998}. We used the coefficients listed in \cite{Adam2017a}, together with their \textit{XMM-Newton} X-ray spectroscopic temperature map, to convert from surface brightness to Compton parameter as a function of temperature.

The Compton parameter is related to the electron pressure $P_e$, via integration along the line-of-sight $\ell$, such that
\begin{equation}
y = \frac{\sigma_{\rm T}}{m_{\rm e} c^2} \int P_{\rm e} d \ell.
\label{eq:compton_parameter_def}
\end{equation}
The parameter $\sigma_{\rm T}$ is the Thomson cross section, $m_{\rm e}$ is the electron rest mass, and $c$ is the speed of light.

In \texttt{PITSZI}, we decompose the pressure into a mean radial component and pressure fluctuations such that
\begin{equation}
\begin{split}
P_{\rm e}(x_1,x_2,x_3) & = \bar{P}_{\rm e}(r) + \delta P_{\rm e}(x_1,x_2,x_3) \\
                                      & = \bar{P}_{\rm e}(r) \left(1 + \frac{\delta P_{\rm e}(x_1,x_2,x_3)}{\bar{P}_{\rm e}(r)}\right).
\end{split}
\end{equation}
Here, $x_1$, $x_2$, and $x_3 \equiv \ell$ are the coordinates in the plane of the sky and along the line-of-sight, and $r = \sqrt{x_1^2 + x_2^2  + x_3^2}$ is the three-dimensional radius.

\subsubsection{The mean pressure profile}\label{sec:The_mean_pressure_profile}
The choice of the mean radial distribution is key for fluctuation analyses \citep[see discussions in, e.g.,][]{Romero2023,Dupourque2023}. Here, we discuss how the pressure profile can be modeled in \texttt{PITSZI} and we refer to Appendix~\ref{app:radial_model_implication} for a study of the impact of the choice of the model on the results.

\paragraph{Mean pressure profile}
\texttt{PITSZI} was built taking advantage of the developments made for the \texttt{MINOT} software\footnote{\url{https://github.com/remi-adam/minot}}. As such, it implements the same profile libraries, including: the $\beta$, double-$\beta$, Navarro-Frenk-White (NFW), generalized Navarro-Frenk-White (gNFW), simplified Vikhlinin model (SVM), or any user-defined profile to be interpolated in log-space \citep[see][for details]{Adam2020}. It is also possible to set the pressure profile given the mass and the redshift of the cluster, according to various universal models available from the literature \citep[e.g.,][]{Arnaud2010,PlanckV2013,Ghirardini2019,MelinPratt23}. 

\paragraph{Triaxiality}
The expected shape of dark matter halos is triaxial rather than spherical. Even though the gas should be rounder, its structure is still driven by the dark matter collapse \citep{Limousin2013}. It is therefore mandatory to allow for triaxiality in the radial pressure modeling in order not to confuse pressure fluctuations with deviations from sphericity.

In \texttt{PITSZI}, the shape of the ellipsoid is parametrized via the minor-to-major and intermediate-to-major axis ratio, $q_{\rm min}$ and $q_{\rm int}$, respectively. We define the ellipsoid radius as
\begin{equation}
\xi = \sqrt{\left(\frac{x_1}{q_{\rm min}}\right)^2 + \left(\frac{x_2}{q_{\rm int}}\right)^2 + \left(x_3\right)^2}.
\end{equation}
The pressure profile is then given by $\bar{P}_{\rm e}(r) \longrightarrow \bar{P}_{\rm e}(\xi)$. 

The axes are initially defined so that the major axis is aligned with the line-of-sight, the intermediate axis aligned with the latitude coordinates, and the minor axis aligned with the longitude. The cluster is then rotated according to the three Euler angles, $\phi_1, \phi_2, \phi_3$, to allow for any orientation, using the \texttt{scipy.spatial.transform.Rotation} package. The rotation follows the so-called \textit{ZXZ} sequence, meaning that $\phi_1$ corresponds to a first rotation around the line-of-sight axis ($x_3$), $\phi_2$ corresponds to a second rotation around the longitude axis ($x_1$), and $\phi_3$ corresponds to a third rotation around the line-of-sight axis ($x_3$).

\subsubsection{Description of the pressure fluctuations}

The pressure fluctuations relative to the radial pressure profile $\left(\frac{\delta P_{\rm e}(x_1,x_2,x_3)}{\bar{P}_{\rm e}(r)}\right)$ are modeled as an isotropic random field. This field is described by its power spectrum and a probability density function. While radial evolution of the fluctuations are usually expected \citep{Shaw2010,Battaglia2012,Nelson2014}, this is not currently implemented in \texttt{PITSZI}, which assumes a single pressure fluctuation power spectrum over the whole ICM.

For the time being, only one power spectrum model is available, a cutoff powerlaw function, but any other function can easily be implemented. This 3D model power spectrum, which we use in the following, is parameterized as
\begin{equation}
\mathcal{P}_{\delta P / \bar{P}} (k_{\rm 3D}) = \sigma_{\mathcal{P}}^2 \frac{k_{\rm 3D}^{\alpha} {\rm exp}\left(-\frac{1}{k_{\rm 3D}^2 L_{\rm inj}^2}\right) {\rm exp}\left(-k_{\rm 3D}^2 L_{\rm dis}^2\right)}{\int 4 \pi k_{\rm 3D}^{\alpha+2} {\rm exp}\left(-\frac{1}{k_{\rm 3D}^2 L_{\rm inj}^2}\right) {\rm exp}\left(-k_{\rm 3D}^2 L_{\rm dis}^2\right) dk_{\rm 3D}}.
\end{equation}
Here, the notation $k_{\rm 3D}$ specifies that we consider the power spectrum in three dimensions (contrary to the 2D power spectrum of SZ images discussed later) and $k_{\rm 3D}$ is the norm of the wavenumber $\vec{k} = \left(k_1,k_2,k_3\right)$. The parameter $\sigma_{\mathcal{P}}$ represents the standard deviation of the pressure fluctuations and gives the normalization of the power spectrum\footnote{In practice, the output standard deviation of a pressure fluctuation box obtained by generating a random field that follows this spectrum may be lower than $\sigma_{\mathcal{P}}$ if the scales below and above those sampled by the model are not available.}. The parameter $\alpha$ is the slope of the spectrum in the inertial range. Its canonical value is $\alpha = -11/3$ for a Kolmogorov cascade \citep{Gaspari2014}. The parameters $L_{\rm inj}$ and $L_{\rm dis}$ are the injection scale and the dissipation scale, respectively.

With the power spectrum in hand, it is also useful to consider the unitless characteristic spectrum, defined as 
\begin{equation}
A_{\delta P / \bar{P}} (k_{\rm 3D}) = \sqrt{4 \pi k_{\rm 3D}^3 \mathcal{P}_{\delta P / \bar{P}} (k_{\rm 3D})}.
\label{eq:charac_amplitude}
\end{equation}

The pressure fluctuations are also described given a probability distribution function, for which either a gaussian or a lognormal distribution is available. We use the formalism described in \cite{Greiner2015} to generate lognormal fluctuations that follow the required power spectrum.

\subsubsection{Relation between pressure fluctuation (3D) and SZ surface brightness fluctuations (2D) power spectra}\label{sec:Relation_3D_2D_power_spectra}

The pressure fluctuation power spectrum is related to the SZ surface brightness fluctuation power spectrum by \citep[e.g.,][for the formalism in X-ray and SZ, respectively]{Churazov2012,Khatri2016}
\begin{equation}
\mathcal{P}_{\delta y / \bar{y}} (k_{\rm 2D}) = \int \mathcal{P}_{\delta P / \bar{P}} (k_{\rm 3D}) \ \left|\tilde{W}(k_1, k_2, k_3)\right|^2 \ dk_3.
\label{eq:3dto2dpowerspec}
\end{equation}
The quantity $\mathcal{P}_{\delta y / \bar{y}} (k_{\rm 2D})$ is the power spectrum of the residual SZ image between the radial model $\bar{y}(R)$ and the total SZ image $y(x_1,x_2)$, such that
\begin{equation}
\frac{\delta y}{\bar{y}} \equiv \frac{y(x_1,x_2) - \bar{y}(R)}{\bar{y}(R)},
\end{equation}
with $R = \sqrt{x_1^2+x_2^2}$ the projected radius and $k_{\rm 2D}$ the norm of the wavenumber in the plane of the sky. The quantity $\tilde{W}(k_1, k_2, k_3)$ is the Fourier transform of the window function given by
\begin{equation}
W(x_1,x_2,x_3) = \frac{\sigma_{\rm T}}{m_{\rm e} c^2} \frac{\bar{P_{\rm e}}(r)}{\bar{y}(R)}.
\end{equation}

In practice, the window function quickly drops for $k_3$ above a given cutoff, such that Equation~\ref{eq:3dto2dpowerspec} is well approximated by
\begin{equation}
\mathcal{P}_{\delta y / \bar{y}} (k_{\rm 2D}) \simeq \mathcal{P}_{\delta P / \bar{P}} (k_{\rm 3D}) \int \left|\tilde{W}(k_1, k_2, k_3)\right|^2 \ dk_3.
\label{eq:3dto2dpowerspec_approx}
\end{equation} 
Using this simplification implies that the map
\begin{equation}
C_{3D \rightarrow 2D} = \int \left|\tilde{W}(k_1, k_2, k_3)\right|^2 \ dk_3
\end{equation}
provides a simple normalization to convert the 3D to 2D power spectrum. In practice, one needs to average the conversion factor over the effective area over which the power spectrum is computed to obtain
\begin{equation}
C_{3D \rightarrow 2D}^{({\rm eff})} = \frac{\sum_{\rm map \ pixels} \omega(x_1,x_2)^2 C_{3D \rightarrow 2D}(x_1,x_2)}{\sum_{\rm map \ pixels} \omega(x_1,x_2)^2},
\end{equation}
where $\omega(x_1,x_2)$ is a weight map (e.g. data mask, region of interest, or any weighting function that is applied to the data prior computing the power spectrum ; see Section~\ref{sec:accounting_for_weighting}). In the end, Equation~\ref{eq:3dto2dpowerspec} becomes
\begin{equation}
\mathcal{P}_{\delta y / \bar{y}} (k_{\rm 2D}) = C_{3D \rightarrow 2D}^{({\rm eff})} \times \mathcal{P}_{\delta P / \bar{P}} (k_{\rm 3D}).
\label{eq:projection_only_modeling}
\end{equation}
This approximation was introduced by \cite{Churazov2012}, who studied the X-ray surface brightness fluctuations in the core of the Coma cluster with XMM-Newton and Chandra. However, it may significantly break at large scales, especially for 3D power spectra with low power on large scales. Additionally, the approximation holds better for flatter pressure profiles since the window function in this case would be sharper (see also \citealt{Clerc2019} for more insights). The limits of this approximation are quantified in Appendix~\ref{app:precision_of_the_model}, where we see that it is in fact fairly good in the vast majority of practical cases. In the present analysis, it implies a bias on the power spectrum model that remains within about 10\% over the range of scales that we probe.

\subsection{Power spectrum measurement and modeling accounting for the instrumental response, data weighting and noise}

After calibration, the observed Compton parameter map obtained from SZ imaging instruments can usually be expressed as
\begin{equation}
y_{\rm observed} = H \circledast B \circledast y_{\rm true} + n,
\end{equation}
where $y_{\rm true}$ is the true incoming signal. The quantity $B$ represents the beam smoothing induced by the limited telescope angular resolution and $H$ is the transfer function induced by the data reduction that usually filters the scales larger than the field of view \citep{Adam2015,Romero2017,Ruppin2018}. The term $n$ is the noise. The map used to extract the 2D SZ power spectrum is defined as 
\begin{equation}
S = \omega \times \frac{y_{\rm observed} - y^{(\rm num)}_{\rm radial \ model}}{y^{(\rm den)}_{\rm radial \ model}},
\label{eq:signal_map_def}
\end{equation}
with $\omega$ the weight map. The radial component model in the numerator and denominator, $y^{(\rm num, \ den)}_{\rm model}$, may slightly differ (see below). 

Let us now consider the measurement of the power spectrum itself and its modeling in the presence of these instrumental effects. This includes beam smoothing and transfer function filtering due to the mapmaking procedure, data masking or weighting, and noise.

\subsubsection{Instrumental response function}

Leaving the weight and the noise aside for the moment, in Fourier space, application of the instrument response function translates into
\begin{equation}
\mathcal{P}_{\rm observed} (k_{\rm 2D}) \simeq B(k_{\rm 2D})^2 H(k_{\rm 2D})^2 \mathcal{P}_{\delta y / \bar{y}} (k_{\rm 2D}).
\label{eq:irfs_application}
\end{equation}
Equation~\ref{eq:irfs_application} is not a strict equality due the fact that the convolution of the instrument response with the observable $S$ is not the same as the ratio of the convolved numerator and denominator. While $y^{(\rm num)}$ accounts for the instrumental response to match the observed data, we therefore define $y^{(\rm den)}$ such that it does not account for the transfer function, but is smoothed by the beam. This implies that deviations from equality only appears on scales below the beam cutoff, where the information on fluctuations is lost anyway.

In the case where the cross spectra between two maps $a$ and $b$ are used, the beam and transfer function become $B^2(k_{\rm 2D}) \rightarrow B^{(a)}(k_{\rm 2D}) \times B^{(b)}(k_{\rm 2D})$ and $H^2(k_{\rm 2D}) \rightarrow H^{(a)}(k_{\rm 2D}) \times H^{(b)}(k_{\rm 2D})$, respectively. 

For a Gaussian beam with ${\rm FWHM} = 2 \sqrt{2 \ {\rm ln}(2)} \sigma_{\rm beam}$, the transmission reads
\begin{equation}
B(k_{\rm 2D}) = {\rm exp}\left(-2 \pi^2 k_{\rm 2D}^2 \sigma_{\rm beam}^2\right).
\end{equation}
The exact shape of the transfer function will depend on the details of the data reduction and shall be provided as part of the data products for any quantitative analysis.

\subsubsection{Data weighting}\label{sec:accounting_for_weighting}
In practice, the $\delta y / \bar{y}$ image is not used to extract the power spectrum directly. Indeed, one needs to account for possible masking, or more generally data weighting. This may arise due to the presence of contaminant point sources that cannot be subtracted perfectly, because we might want to extract the power spectrum over a given region of interest (e.g., within $R_{500}$), or in the case where data weighting is required (e.g., to account for noise inhomogeneity). In \texttt{PITSZI}, we implemented the methodology developed by \cite{Ponthieu2011} for the \texttt{POKER} software in order to account for data masking and weighting in the power spectrum modeling.

Assuming the SZ fluctuation image to be weighted as $S \equiv \omega \times  \delta y / \bar{y}$, its pseudo-Fourier coefficients, for each mode $m_1,m_2$, are given by
\begin{equation}
\tilde{S}_{m_1,m_2} = \sum_{\mu_1,\mu_2} \omega_{\mu_1, \mu_2}  \left(\frac{\delta y}{\bar{y}}\right)_{\mu_1, \mu_2} e^{-2 i \pi \left(\mu_1 m_1 / N_1 + \mu_2 m_2/N_2\right)},
\end{equation}
with $N_1,N_2$ the number of pixels along latitude and longitude, indexed $\mu_1$ and $\mu_2$, respectively. We can show that \citep[][their appendix A, albeit with different Fourier transform normalization]{Ponthieu2011} 
\begin{equation}
\tilde{S}_{m_1,m_2} = \sum_{m_1^{\prime}, m_2^{\prime}} \tilde{S}_{m_1^{\prime}, m_2^{\prime}}^{\rm unweighted} \frac{K_{m_1, m_1^{\prime}}^{m_2, m_2^{\prime}}}{N_1 N_2} ,
\label{eq:Kmn_application}
\end{equation}
with $\tilde{S}_{m_1^{\prime}, m_2^{\prime}}^{\rm unweighted}$ the pseudo-Fourier coefficients of the unweighted image $\delta y / \bar{y}$ and $K_{m_1, m_1^{\prime}}^{m_2, m_2^{\prime}}$ a matrix that depends on the pseudo-Fourier coefficients of the weight map. The power spectrum of the weighted image is given by $\mathcal{P}_{m_1,m_2} = \left|\tilde{S}_{m_1,m_2}\right|^2$ for each mode $m_1,m_2$ and $\mathcal{P}(k_{\rm 2D})$ is obtained by binning $\mathcal{P}_{m_1,m_2}$ in $k_{\rm 2D}$ assuming isotropy. We refer to \cite{Ponthieu2011} for more insight into the mathematical details.

The comparison between the power spectrum of weighted data and a model of SZ fluctuations should therefore either convolve the model to account for the effect of the weights (using Equation~\ref{eq:Kmn_application} together with binning), or deconvolve the data from the weights. Moreover, it is advisable to embed the observed sky patch into a larger patch and pad it with zeros to prevent aliasing from scales larger than the observed sky. The observation patch should be apodized to minimize large-scale aliasing. Other methods have been used in the literature to estimate the power spectrum in this context. In particular, most results have used the method from \cite{Arevalo2012}, which differs from ours. Appendix \ref{app:precision_of_the_model} quantifies the underlying uncertainties.

\subsubsection{Noise}\label{sec:noise}
The contribution from the noise is obtained by applying the same process to noise Monte Carlo realizations. In this case, we use the image given by 
\begin{equation}
S_{{\rm noise}, i} = \omega \times \frac{n_i}{y^{(\rm den)}_{\rm radial \ model}},
\label{eq:noise_MC}
\end{equation}
for each Monte Carlo $i$, to obtain the power spectra $\mathcal{P}_{{\rm noise}, i} (k_{\rm 2D})$. The mean across all the Monte Carlo realizations provides the expected noise bias $\mathcal{P}_{\rm noise} (k_{\rm 2D}) = \left<\mathcal{P}_{{\rm noise}, i} (k_{\rm 2D})\right>$ and the noise covariance matrix is given by $C_{\rm noise} = \left< \left(\mathcal{P}_{{\rm noise}, i} (k_{\rm 2D}) - \mathcal{P}_{\rm noise} (k_{\rm 2D})\right) \left(\mathcal{P}_{{\rm noise}, i} (k_{\rm 2D}) - \mathcal{P}_{\rm noise} (k_{\rm 2D})\right)\right>$. It will be an essential ingredient for the likelihood estimation and power spectrum estimation (Section~\ref{sec:inference}). Note that model uncertainties are not accounted in Equation~\ref{eq:noise_MC} because they will be treated as a systematic effect.

When two independent data (sub)sets are used to compute the power spectrum, the noise is estimated accordingly. We thus use the cross spectra between two independent noise realizations $S_{{\rm noise}, i}^{(a)}$ and $S_{{\rm noise}, i}^{(b)}$. In this case, the mean noise power spectrum is expected to be zero.

The exact same procedure allows us to extract the contribution from the CIB, to obtain $\mathcal{P}_{\rm CIB} (k_{\rm 2D})$ and $C_{\rm CIB}$, or any astrophysical contaminant in general.

\subsection{Summary, discussion and example}\label{sec:Summary_discussion_example}
\begin{figure*}
        \centering
        \includegraphics[width=0.99\textwidth]{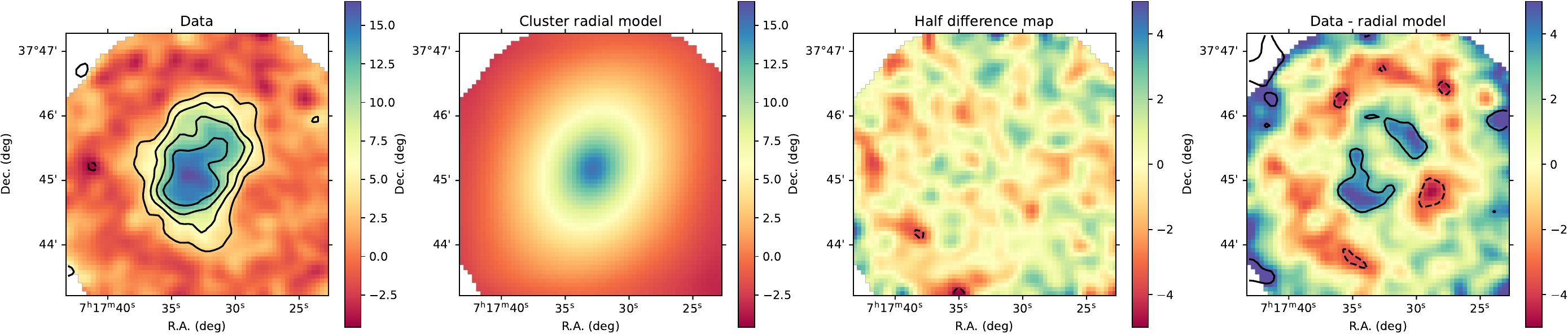}
        \caption{NIKA Compton parameter ($y$) data. From left to right, we show the NIKA MACS~J0717.5+3705 input data, its best-fit radial model map, a jackknife map (half difference between two equal data subsets) and the residual between the data and the radial model, $\delta y$. The data were smoothed with a 15 arcsec gaussian kernel (FWHM) for visual purposes. The contours are showing the S/N in units of $3 \sigma$.
        The comparison with mock data can be observed in Figure~\ref{fig:example_images_models}.
        }
\label{fig:example_images_data}
\end{figure*}

\begin{figure*}
        \centering
        \includegraphics[width=0.99\textwidth]{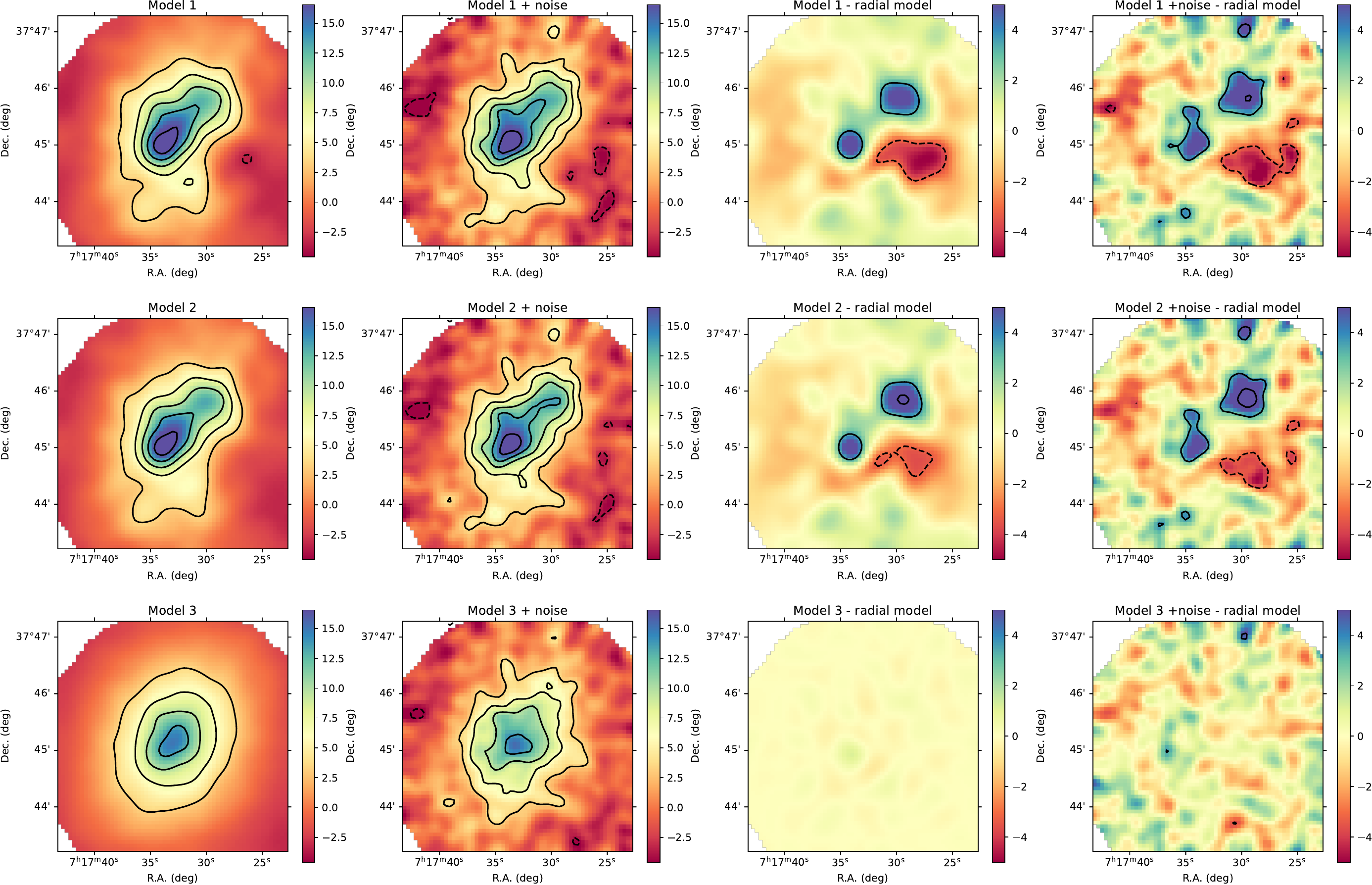}
        \caption{Simulated SZ mocks to be compared with the NIKA data (Figure~\ref{fig:example_images_data}). The smoothing kernel and the contours are the same as Figure~\ref{fig:example_images_data}.
        From left to right, we show a noiseless SZ realization, the same image with noise, the noiseless residual between the mock and the radial model, and the residual including noise.
        The first, second and third rows correspond to model 1, model 2 and model 3, respectively. See Table~\ref{tab:test_model} for the definition of the models.
        }
\label{fig:example_images_models}
\end{figure*}

\begin{figure*}
        \centering
        \includegraphics[width=0.33\textwidth]{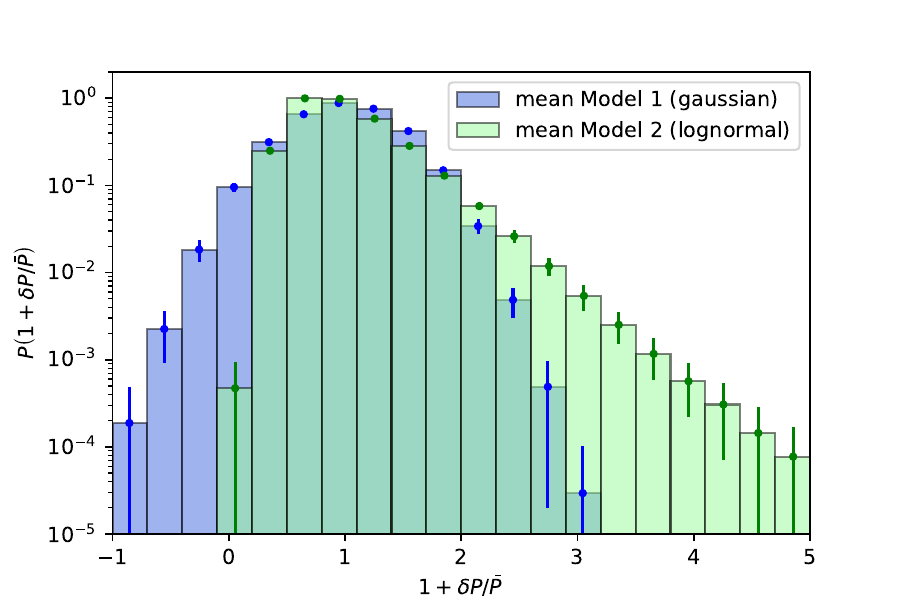}
        \includegraphics[width=0.33\textwidth]{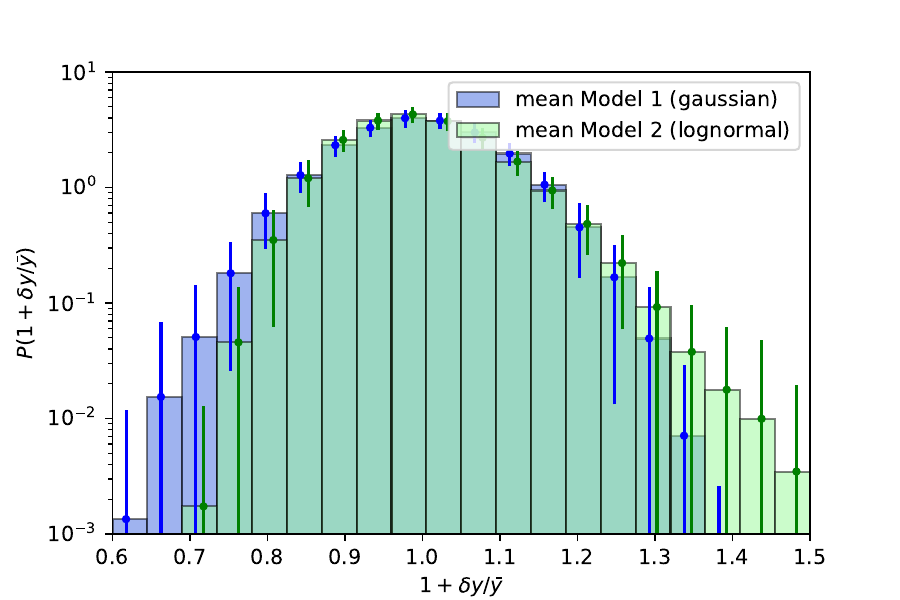}
        \includegraphics[width=0.33\textwidth]{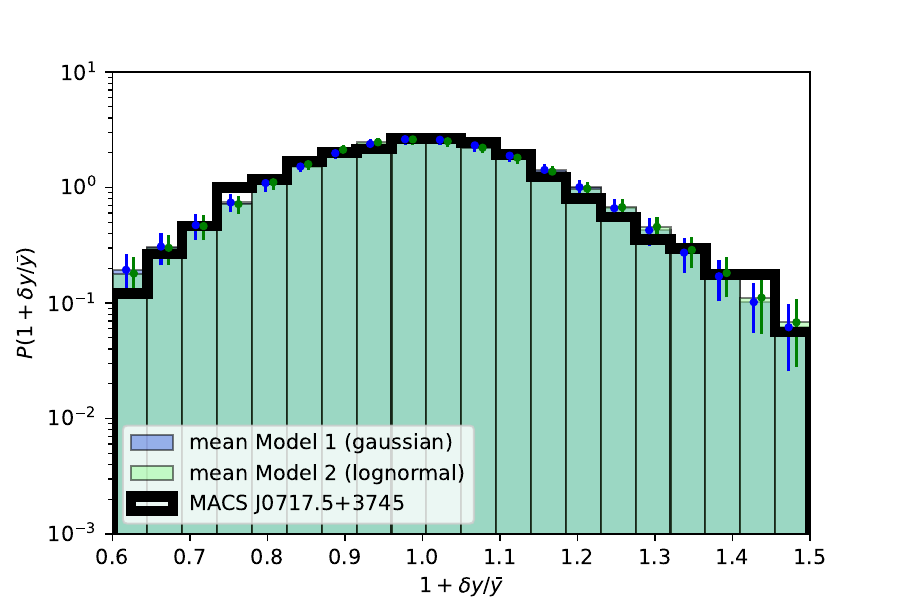}
        \caption{Comparison between the pressure and SZ fluctuations 1D probability distributions for model 1 (gaussian) and model 2 (lognormal), and MACS~J0717.5+3745 data.
        {\bf Left}: probability density function of the pressure fluctuations $1 + \delta P / \bar{P}$ for model 1 (gaussian fluctuations) and model 2 (lognormal fluctuations).
        {\bf Middle}: probability density function of the SZ fluctuations $1+\frac{\delta y}{\bar{y}}$ for model 1 and model 2, without noise.
        {\bf Right}: probability density function of the SZ fluctuations $1+\frac{\delta y}{\bar{y}}$ for model 1 and model 2, including noise, and comparison with MACS~J0717.5+3745 NIKA data. The data and model were smoothed with a 10 arcsec FWHM gaussian kernel to reduce the noise.
        The mean distributions and error bars were obtained from the mean and standard deviations, respectively, of 100 Monte Carlo realizations.
        }
\label{fig:example_statistics}
\end{figure*}

\begin{figure*}
        \centering
        \includegraphics[width=0.99\textwidth]{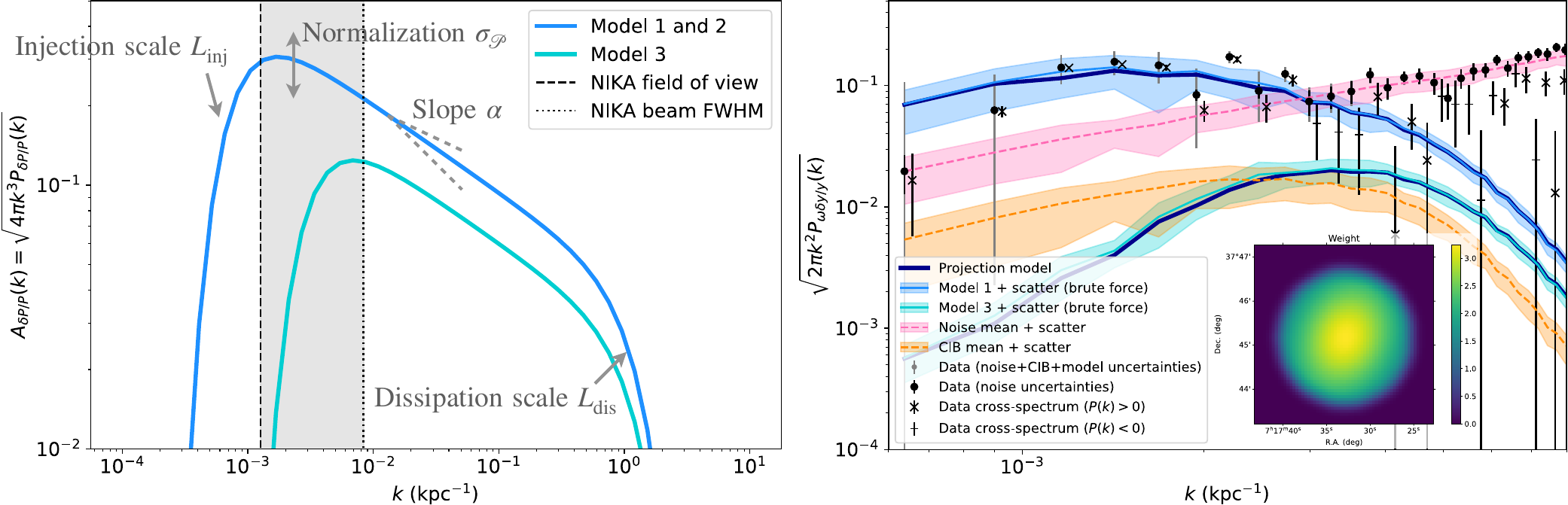}
        \caption{
        {\bf Left:} input 3D pressure fluctuations power spectra for model 1, 2, and 3. The grey area roughly corresponds to the observable modes with the NIKA data.
        {\bf Right:} 2D SZ fluctuations power spectra for MACS~J0717.5+3745 NIKA data, the noise contribution, the CIB contribution, and models 1 and 3 including the projection and brute force approaches. The shaded area corresponds to the measured standard deviation of the respective spectra. For the data, error bars indicate the contribution from the noise and CIB only in black, and account for the intrinsic scatter of model 1 in grey. We show both the auto spectrum of the full data, and the cross spectra of two independent data subsets. For the later, the power spectrum may be negative, as indicate in the legend. The uncertainties due to the noise are $\sqrt{2}$ larger for the cross spectrum than for the auto spectrum.
        }
\label{fig:example_pk}
\end{figure*}

In \texttt{PITSZI}, the power spectrum of the data is computed as the modulus squared of the fast Fourier transform of the quantity $S$ defined in Equation~\ref{eq:signal_map_def}, which is then binned in $k_{\rm 2D}$ to obtain $\mathcal{P}_{\rm observed} (k_{\rm 2D})$. It is also possible to use the cross spectra between two independent data (sub)sets, $S^{(a)}$ and $S^{(b)}$, if available. The weight map accounts for the region of interest and discarded pixels. In practice, the weights are somewhat arbitrary. For instance, one might set $\omega$ to $y^{(\rm den)}_{\rm model}$ to avoid a dramatic increase in the noise with radius. However, this will be at the cost of obtaining a power spectrum estimate that will be weighted towards the cluster center.

To obtain a mock or an actual model power spectrum that can be compared to the data, one must account for the instrument response function and the weighting scheme. Starting from a 3D pressure fluctuation power spectrum model, \texttt{PITSZI} implements two different approaches. The projection approach computes the model as follows: 1) Equation~\ref{eq:projection_only_modeling} is used to compute the theoretical 2D power spectrum; 2) Equation~\ref{eq:irfs_application} is applied to account for the instrument response function; 3) the power spectrum is projected in $k$-space given the sampled mode $m_1,m_2$ and Equation~\ref{eq:Kmn_application} is applied to account for data weighting; 4) the power spectrum is binned in $k_{\rm 2D}$ to obtain $\mathcal{P}_{\rm ICM} (k_{\rm 2D})$. Alternatively, a brute force approach may be used in which full SZ image Monte Carlo simulations are generated and their power spectra calculated as for the data. This is done by generating mock pressure cubes including the radial profile and the fluctuations; obtaining SZ images via Equation~\ref{eq:compton_parameter_def}; convolving the image with the instrument response function; applying the weights; and finally extracting the 2D power spectrum. As in the case of the noise (Section~\ref{sec:noise}), the statistics of the power spectra obtained in this manner allow us to determine the mean power spectrum $\mathcal{P}_{\rm ICM} (k_{\rm 2D}) = \left<\mathcal{P}_{{\rm Monte \ Carlo}, i} (k_{\rm 2D})\right>$ and its covariance matrix $C_{\rm ICM}$. The brute force approach is significantly more computationally expensive than the projection approach, and is not used to fit the data.

To illustrate the methodology discussed in this Section, we now compare the MACS~J0717.5+3745 NIKA data with three mock SZ maps constructed from different pressure fluctuation models with \texttt{PITSZI} (model~1, model~2 and model~3, see Table~\ref{tab:test_model}). These are designed to qualitatively mimic the NIKA data, but we refer to Section~\ref{sec:inference} for a full discussion of model inference. The radial model is the same for the three cases, and is obtained by fitting an elliptical gNFW model to the NIKA data
\begin{equation}
\bar{P}(r) = \frac{P_0}{\left(\frac{r}{r_p}\right)^{c} \left(1 + \left(\frac{r}{r_p}\right)^{a}\right)^{\frac{b-c}{a}}}.
\label{eq:gnfw}
\end{equation}
The slope parameters $a$, $b$ and $c$ are fixed to the morphologically disturbed model of \cite{Arnaud2010}, the center is free to vary, and the mass is fit via a single normalization parameter that also defines the scale radius\footnote{In Equation~\ref{eq:gnfw}, the characteristic radius $r_p = R_{500}/c_{500}$ and the normalization $P_0$ can be related to the mass assuming that clusters follow the same scaled pressure profile. See \cite{Arnaud2010} for details.}. The first fluctuation model is built with $\sigma_{\mathcal{P}} = 0.5$, $L_{\rm inj} = 1$~Mpc, and $\alpha=-11/3$, with Gaussian pressure fluctuations. The second model has the same power spectrum parameters but with lognormal pressure fluctuations (and we use the same seed to generate random fluctuations). The third model is built with $\sigma_{\mathcal{P}} = 0.2$, $L_{\rm inj} = 250$ kpc, and $\alpha=-11/3$, with Gaussian pressure fluctuations. In all cases, $L_{\rm dis} = 1$ kpc, but the exact value of this parameter is irrelevant here because it lies well below the instrumental resolution.
\begin{table}[h]
\caption{\footnotesize{Summary of the test model properties used for illustration.}}
\begin{center}
\resizebox{0.4\textwidth}{!} {
\begin{tabular}{c|cccc}
\hline
\hline
   & $\sigma_{\mathcal{P}}$ & $L_{\rm inj}$ & $\alpha$ & Statistics \\
\hline
Test model 1 & 0.5 & 1 Mpc    & -11/3 & Gaussian \\
Test model 2 & 0.5 & 1 Mpc    & -11/3 & Lognormal \\
Test model 3 & 0.2 & 250 kpc  & -11/3 & Gaussian \\
\hline
\end{tabular}
}
\end{center}
\label{tab:test_model}
\end{table}

In Figure~\ref{fig:example_images_data}, we can see that, given the elliptical radial model, SZ fluctuations are observed by eye in the NIKA residual map. The half-difference (Jackknife) map gives a visual comparison of the noise contribution to the data. In Figure~\ref{fig:example_images_models}, the SZ mocks derived from model~1 and~2 visually resemble the data, at least qualitatively. When including the noise Monte Carlo, it would be nearly impossible to distinguish mock images from true data. This indicates that our description of the ICM as a radial model plus isotropic pressure perturbations mimics the real data well, at least given the sensitivity and the scales that are probed with NIKA. We defer discussion of the limitations of the modeling approach to Section~\ref{sec:physics}. Contrary to models~1 and~2, we can already observe that model~3 presents too few fluctuations and its residual is nearly consistent with the half difference map of the true data. Visually, models~1 and~2 are nearly identical, and look statistically similar. 

In Figure~\ref{fig:example_statistics}, we show that indeed, the fluctuation statistics (lognormal or Gaussian) cannot be distinguished for the pressure fluctuation models~1 and~2. While the pressure fluctuation distributions present significant differences, when seen as SZ fluctuations models~1 and~2 are nearly compatible, even without including noise in the data, owing to projection effects. The Gaussian versus lognormal models become significantly different only when the fluctuation amplitude is significantly increased. When including noise, it is not possible to distinguish the two within uncertainties. The fluctuations obtained for MACS~J0717.5+3745 are compatible with those expected in both models~1 and~2. 

In Figure~\ref{fig:example_pk}, we show the input power spectrum of the pressure fluctuations for the different models. The scales accessible with NIKA roughly correspond to the shaded area, since the signal is progressively filtered for scales larger than the field of view, and on scales smaller than the beam. We thus expect sensitivity to the normalization, the injection scale, and possibly the slope if a sufficient signal-to-noise ratio is available. The dissipation scales, on the other hand, is expected to be far beyond the reach of NIKA in the case of MACS~J0717.5+3745. We also compute the power spectrum of the corresponding SZ fluctuations with our different methods and include the different contributions to the data. Here, the weight map, shown in the bottom right of the right panel, was defined so that pixels beyond $\theta_{500}/2 \simeq 2$ arcmin were masked and the ICM was weighted (i.e. the  $\omega$ term) according to the smooth radial model (with inner slope set to zero) and smoothed with a 20 arcsec FWHM Gaussian kernel to minimize aliasing. We can see that the projection approach matches very well the mean brute force approach, indicating that the approximations discussed in this Section are robust over the full range of scales. The contribution of the CIB is expected to be lower than the noise, but it is still non-negligible. While model~1 is expected to be well detected, and in fact match the data relatively well, model~3 is below the noise and the CIB contributions. Model~2 is not represented because it is very similar to model~1. The shaded regions provide the standard deviation of the different power spectra, including that of the pressure fluctuation model, which arise from the fact that they intrinsically correspond to stochastic processes. The data show both the auto spectrum of the full data and the cross spectrum of two independent equivalent data subsets. In both cases, a clear excess associated with the injected signal is observed in the range $k \sim 10^{-3} - 3 \times 10^{-3}$ kpc$^{-1}$. The auto-spectrum converges to the expected noise mean at high $k$, where it dominates, while the cross-spectrum is scattered around zero. The auto- and cross-spectra agree very well on scales where the signal dominates. The noise contribution dominates the uncertainties at high $k$, while the intrinsic model scatter dominates at low $k$. In the case of MACS~J0717.5+3745, the CIB contribution to the uncertainties is subdominant at all scales, although its mean contribution is not negligible.

\section{Pressure profile and pressure fluctuation inference}\label{sec:inference}

In Section~\ref{sec:modeling}, we  discussed how to compute the SZ fluctuations 2D power spectrum, and how to derive the corresponding model given a 3D pressure fluctuation power spectrum and radial model. Here we will discuss the methodology implemented in \texttt{PITSZI} to infer constraints on the radial model and the pressure fluctuation power spectrum, and then apply it to MACS~J0717.5+3745 data.

\subsection{Pressure profile}\label{sec:pressure_profile}
\begin{table*}[h]
\caption{\footnotesize{Best-fit parameters obtained for  the different radial models. Uncertainties on the radial model parameters are negligible compared to the systematic effects associated with the choice of the model, and are thus omitted.}}
\begin{center}
\resizebox{0.8\textwidth}{!} {
\begin{tabular}{c|cccccccccc}
\hline
\hline
   & R.A.   & Dec.  & $q_{\rm int, \ maj}$ & $\phi_3$ & $M_{500}$                      & $P_0$                & $r_p$ & $a$ & $b$ & $c$ \\
-- & (deg) & (deg) & (--)                             & (deg)      & ($10^{14}$ M$_{\odot}$) & ($10^{-2}$ keV cm$^{-3}$) & (kpc)  & (--) & (--) & (--) \\
\hline
RM$_1$ & 109.3806 & 37.7583  & 1      & --  & 18.1 & -- & -- & -- & -- & -- \\
RM$_2$ & 109.3872 & 37.7530  & 1      & --  & 21.7 & -- & -- & -- & -- & -- \\
RM$_3$ & 109.3878 & 37.7528 & 0.77 & 23 & 26.5 & -- & -- & -- & -- & -- \\
RM$_4$ & 109.3874 & 37.7525 & 0.74 & 24 & --  & 11.7 & 852 & -- & -- & -- \\
RM$_5$ & 109.3870 & 37.7543 & 0.71 & 21 & --  & 10.5 & 463 & 6.6 & 7.4 & 0.06 \\
\hline
\end{tabular}
}
\end{center}
{\footnotesize {\bf Notes.} 
$^{\dagger}$the results shown here were obtained using the non-linear least square method.
}
\label{tab:inference_radial_results}
\end{table*}

\begin{figure*}
        \centering
        \includegraphics[width=\textwidth]{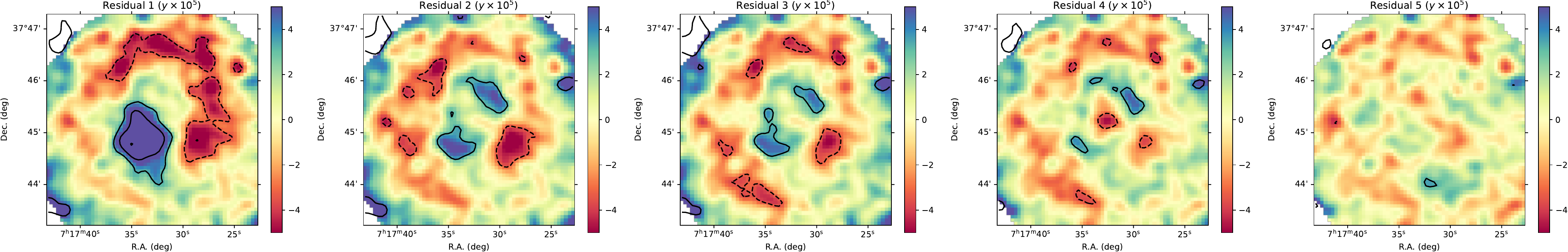}
        \caption{Residual SZ maps between the data and the radial models under consideration. Contours are multiples of $3 \sigma$, and the maps have been smoothed with a 15 arcsec FWHM gaussian kernel for visual purposes.}
\label{fig:radial_model_residuals}
\end{figure*}

The fitting approach used to constrain the radial pressure profile consists of comparing the model, including projection effects and instrumental effects as described in Section~\ref{sec:Pressure_distribution_and_projection}, to the data. While the ICM triaxiality can be modeled in three dimensions, SZ data alone cannot constrain the ICM shape along the line-of-sight \citep[see, e.g.,][for such an analysis]{Kim2024}. Therefore, in the fitting process, only a single axis ratio (assuming $q_{\rm int} = q_{\rm min}$) and the rotation angle $\phi_3$ (assuming $\phi_1 = 0$ and $\phi_2 = 90$ deg) are fit. This corresponds to a fit of the ellipticity as projected onto the plane of the sky. At this stage, \texttt{PITSZI} also accounts for the map zero level as a nuisance parameter, since it is generally unconstrained from SZ observations. In the end, \texttt{PITSZI} allows for the fit of any pressure profile model parameters, the cluster center, the cluster ellipticity, and the map zero level.

Two different algorithms are implemented to constrain the model:  sampling of the parameter space via Markov Chain Monte Carlo (MCMC) using the \texttt{emcee} package \citep{Foreman2013}, and the use of a non-linear least square fit via the \texttt{scipy.optimize.curve\_fit} package\footnote{All parameters may be fixed, sharp limits can be imposed, and  in the case of MCMC, Gaussian priors can be used.}. The MCMC fit uses a Gaussian log likelihood computed as
\begin{equation}
{\rm ln} \ \mathcal{L} \propto \sum_{m,n} \left(D - M\right)_m \left(C^{-1}\right)_{m,n} \left(D-M\right)_n
\end{equation}
where $D$ represents the unmasked data map, $M$ the radial model, and $C$ the noise covariance matrix. In the present paper, we neglect pixel-to-pixel correlations and only account for the noise standard deviation in the fit. This is because obtaining the full noise covariance matrix is computationally expensive, and because its use essentially affects the posterior uncertainties on the parameters which are dominated by systematic uncertainties associated to the choice of radial model. In the case where the posterior is well approximated by a multivariate Gaussian distribution, the MCMC and non-linear least square methods give fully consistent results. In the present analysis, we will neglect the statistical uncertainties associated with the fit parameters of the radial model, because they are much smaller than the systematic uncertainties induced by the choice of the radial model.

Indeed, the choice of radial model is known to be key for surface brightness fluctuation analysis \citep[e.g.,][]{Dupourque2023,Romero2023} and is responsible for most of the systematic uncertainties. Therefore, we explore the stability of the results by using the following radial models (RM):
\begin{enumerate}
\item RM$_1$: center fixed to the X-ray peak (roughly consistent with X-ray barycenter) and a spherical profile. The shape of the pressure profile is fixed to the morphologically disturbed model from \cite{Arnaud2010}, and the mass $M_{500}$ is the only free parameter of the pressure profile. The model has two free parameters (including the map zero level);
\item RM$_2$: free center, spherical profile, $M_{500}$ fitted as the normalization of the morphologically disturbed \cite{Arnaud2010} pressure profile ($2 + 1 + 1 = 4$ parameters);
\item RM$_3$: free center, free elliptical shape, $M_{500}$ fitted as the normalization of the \cite{Arnaud2010} morphologically disturbed pressure profile ($2 + 2 + 1 + 1 = 6$ parameters);
\item RM$_4$: free center, free elliptical shape, free gNFW pressure profile normalization $P_0$ and scale radius $r_p$ ($2 + 2 + 2 + 1 = 7$ parameters);
\item RM$_5$: free center, free elliptical shape, free gNFW pressure profile ($2 + 2 + 5 + 1 = 10$ parameters).
\end{enumerate}
In the case of elliptical profiles, we stress that the mass determination is driven by the shape and normalization of the pressure profile along the major axis, so that its meaning should be taken with caution (see Section~\ref{sec:The_mean_pressure_profile}). In the case of RM$_5$, the profile parameters being highly degenerate, we impose a flat prior on the slopes of the gNFW model such that $a \in [0,5]$, $b \in [2,8]$, $c \in [0, 2]$ given the results from \cite{PlanckV2013}. On one hand, if the model is too simplistic, the cluster potential well is not well represented, and we would expect to observe excess residual features unrelated to pressure fluctuations. On the other hand, models that are too complex would be expected to overfit the pressure fluctuations. See Appendix~\ref{app:radial_model_implication} for more discussion. Accordingly, we expect that our baseline model RM$_3$ should provide a fairly good compromise.

In Table~\ref{tab:inference_radial_results}, we give the best fit results obtained for the five models. In Figure~\ref{fig:radial_model_residuals}, we show the SZ residual maps. As expected, the amount of SZ surface brightness residuals diminishes as a function of increasing model complexity. While fitting for the center largely improves the residual, fitting for the ellipticity has a smaller impact. The most complex model is able to account for a substantial fraction of the fluctuations, and no large features are observed visually in the residual. In Appendix~\ref{app:kSZ_systematics}, we also show that attempting to correct for the kSZ signal implies larger residuals, in particular for model RM$_5$.

\subsection{Pressure fluctuations}
\begin{table*}[h]
\caption{\footnotesize{Constraints on the pressure fluctuation parameters. The central value gives the median of the distribution and the uncertainties correspond to the 68\% confidence intervals.}}
\begin{center}
\resizebox{0.6\textwidth}{!} {
\begin{tabular}{cc|cccc}
\hline
\hline
Analysis & Model & $\sigma_{\mathcal{P}}$ & $L_{\rm inj}$ & $A_{\rm noise}$ & $A_{\rm CIB}$ \\
 & & (--) & (kpc) & (--) & (--)  \\ [1mm]
\hline
\hline
Reference$^{\dagger}$ & RM$_1$ & $1.01_{-0.11}^{+0.08}$ & $3128_{-1792}^{+1322}$ & $1.06_{-0.04}^{+0.04}$ & $1.8_{-1.3}^{+1.8}$ \\ [1mm]
Reference$^{\dagger}$ & RM$_2$ & $0.63_{-0.03}^{+0.03}$ & $859_{-62}^{+63}$           & $1.01_{-0.04}^{+0.04}$ & $2.4_{-1.6}^{+1.7}$ \\ [1mm]
Reference$^{\dagger}$ & RM$_3$ & $0.58_{-0.04}^{+0.04}$ & $873_{-75}^{+74}$           & $1.04_{-0.04}^{+0.04}$ & $2.4_{-1.7}^{+1.7}$ \\ [1mm]
Reference$^{\dagger}$ & RM$_4$ & $0.68_{-0.07}^{+0.07}$ & $653_{-78}^{+78}$           & $1.03_{-0.04}^{+0.04}$ & $2.4_{-1.7}^{+1.7}$ \\ [1mm]
Reference$^{\dagger}$ & RM$_5$ & $0.27_{-0.11}^{+0.11}$ & $1010_{-418}^{+410}$     & $1.05_{-0.04}^{+0.04}$ & $2.5_{-1.7}^{+1.7}$ \\ [1mm]
\hline
Cross-spectrum                        & RM$_3$ & $0.64_{-0.04}^{+0.04}$ & $812_{-72}^{+71}$ & 0 (fixed)                        & $2.5_{-1.7}^{+1.7}$ \\ [1mm]
kSZ template subtracted          & RM$_3$ & $0.70_{-0.03}^{+0.03}$ & $852_{-51}^{+49}$ & $1.03_{-0.04}^{+0.04}$ & $2.4_{-1.6}^{+1.7}$ \\ [1mm]
Point sources masked             & RM$_3$ & $0.61_{-0.04}^{+0.04}$ & $851_{-69}^{+72}$ & $1.03_{-0.04}^{+0.04}$ & $2.4_{-1.6}^{+1.7}$ \\ [1mm]
No radial model based weight & RM$_3$ & $0.70_{-0.04}^{+0.04}$ & $822_{-62}^{+62}$ & $0.98_{-0.04}^{+0.04}$ & $2.4_{-1.7}^{+1.7}$ \\ [1mm]
MCMC fit                                  & RM$_3$ & $0.58_{-0.03}^{+0.03}$ & $869_{-60}^{+67}$ & $1.04_{-0.04}^{+0.04}$ & $1.1_{-0.8}^{+1.7}$ \\ [1mm]
\hline
\end{tabular}
}
\end{center}
{\footnotesize {\bf Notes.} 
$^{\dagger}$ auto-spectrum, no kSZ correction, point sources subtracted, radial model based weighting plus 2 arcmin radius region of interest, non-linear least square fit, and projection approach.
}
\label{tab:inference_fluctuation_results}
\end{table*}

\begin{figure*}
        \centering
        \includegraphics[width=0.49\textwidth]{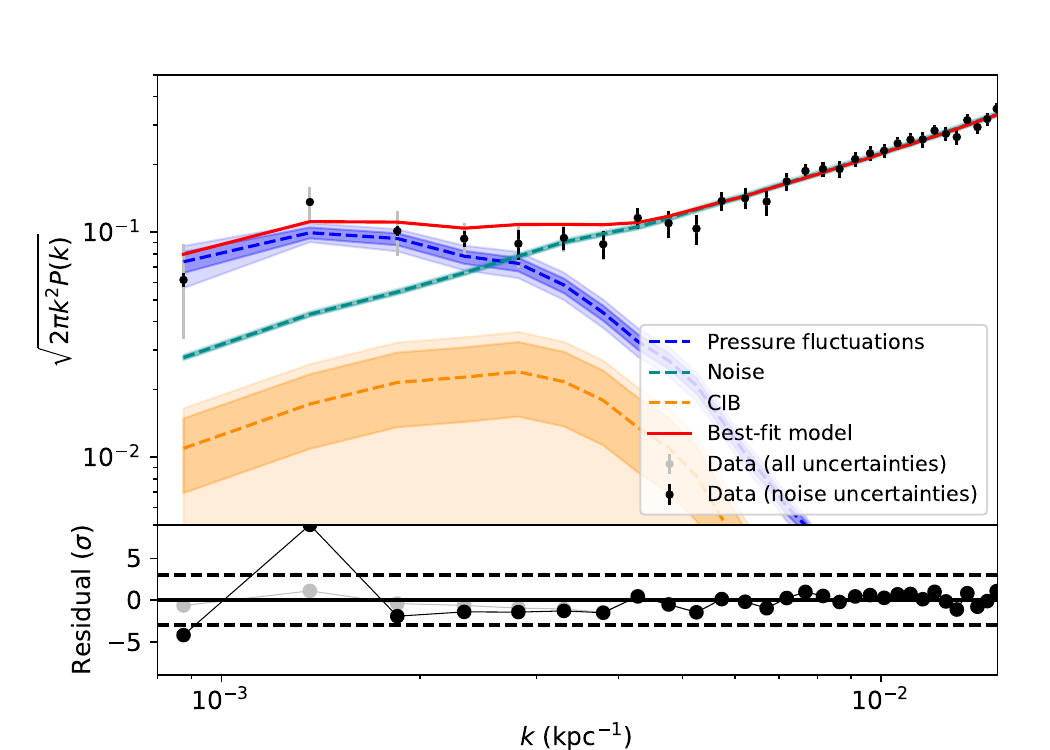}
        \includegraphics[width=0.49\textwidth]{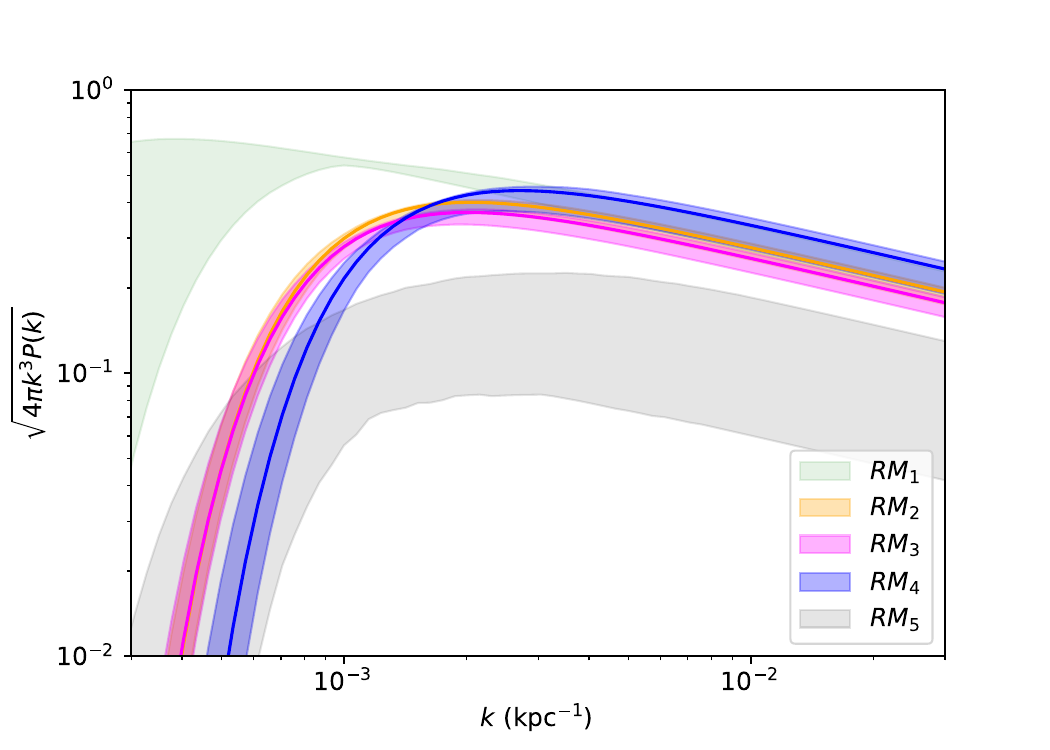}
        \caption{Constraints on the fluctuation model obtained for the reference analysis.
        {\bf Left:} comparison between the data and model for the 2D SZ fluctuation power spectrum, in the case of radial model RM$_3$. The residual shows the offset to the best-fit model relative to the uncertainties, accounting or not for intrinsic model variance (sample variance). Shaded area provide 68\% and 95\% confidence intervals.
        {\bf Right:} constraints on the 3D pressure fluctuation power spectrum for the different radial model, with 68\% confidence interval shown. The sample variance is included in the error budget.
        }
\label{fig:pk}
\end{figure*}

\begin{figure}
        \centering
        \includegraphics[width=0.49\textwidth]{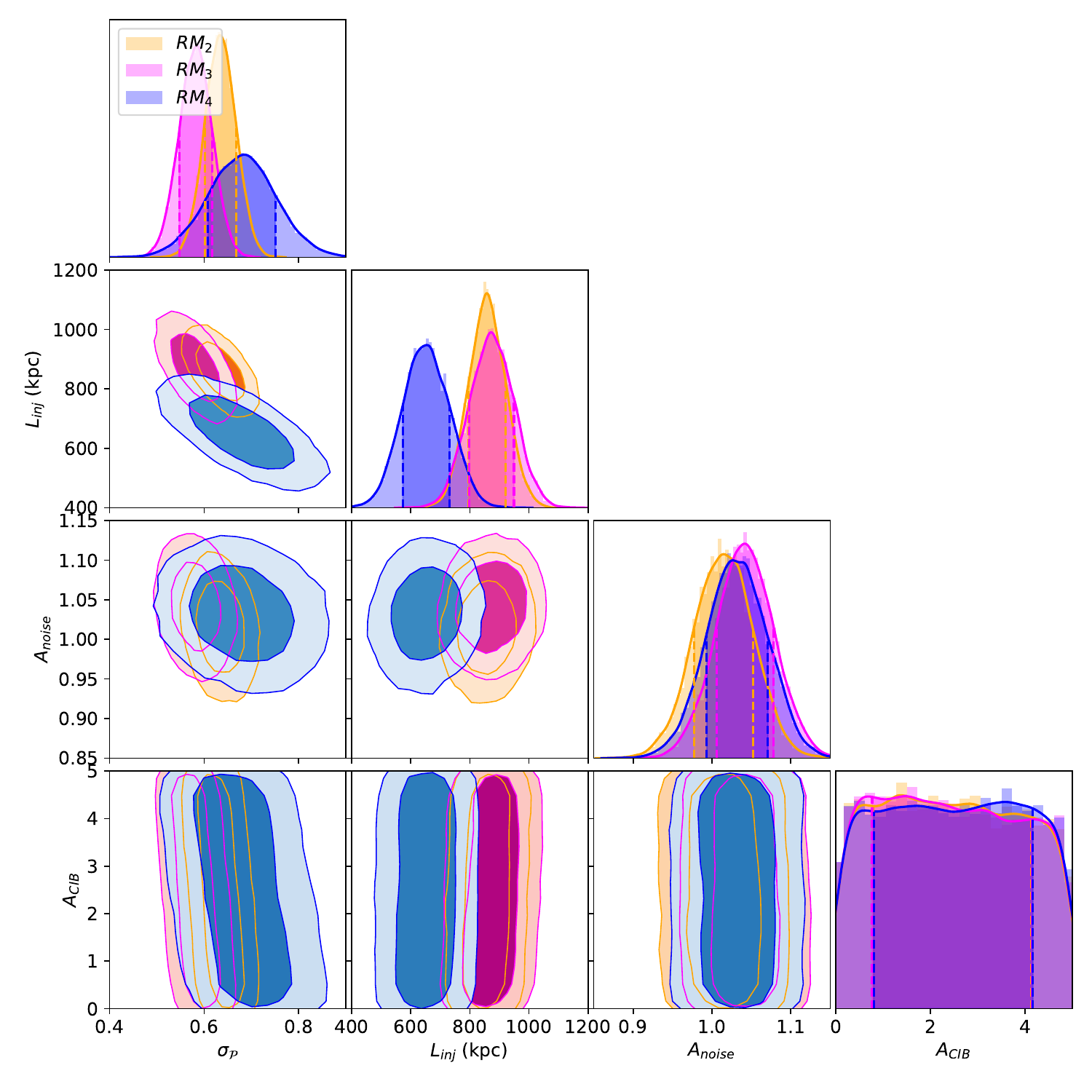}
        \includegraphics[width=0.49\textwidth]{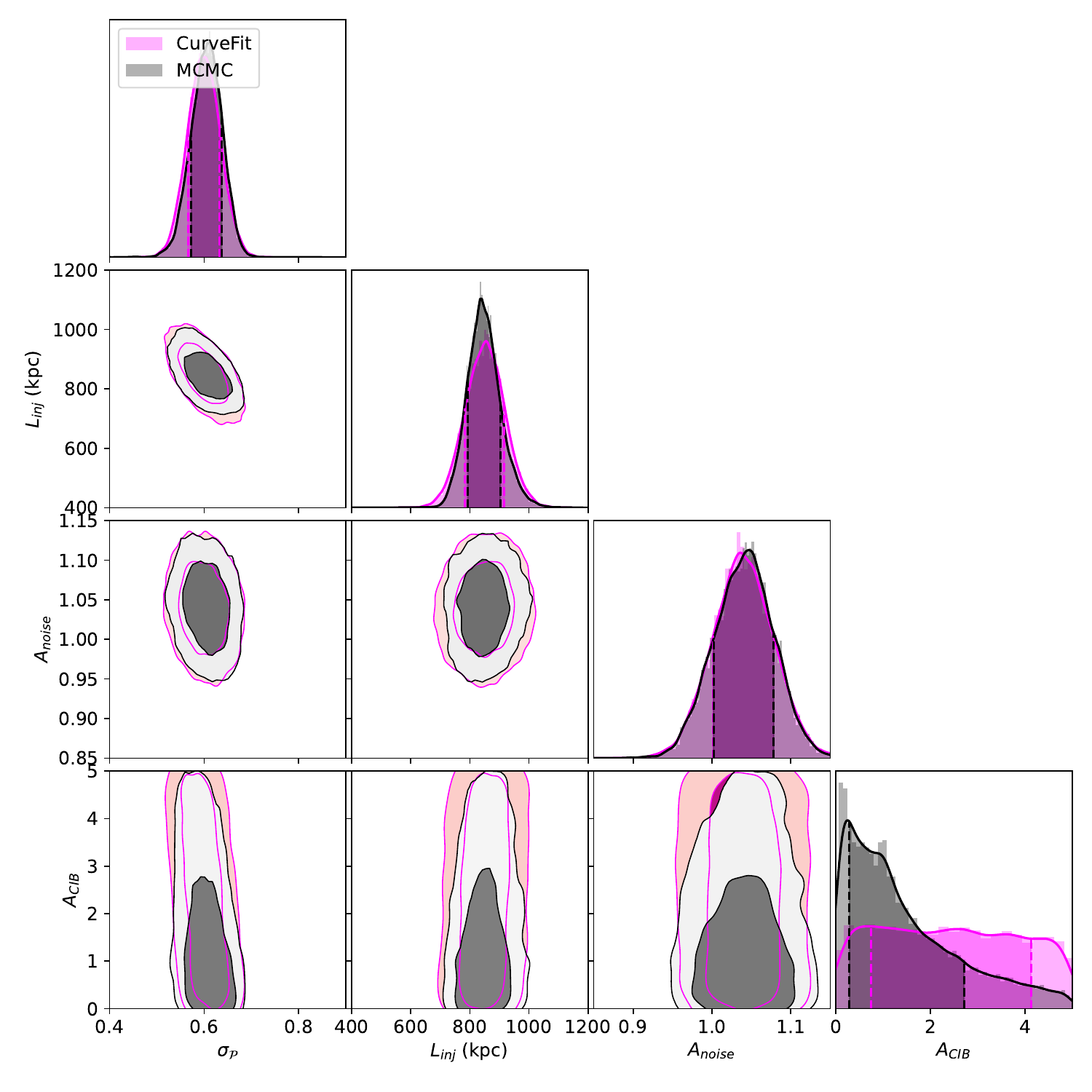}
        \caption{{\bf Top}: Constraints in the parameter space for models RM$_2$, RM$_3$ and RM$_4$. Other extreme models are not shown for clarity. Contours correspond to the 68 and 95\% confidence intervals.
        {\bf Bottom}: comparison between the constraints obtained on RM$_3$ for the full MCMC sampling and the non linear least square approximation methods, in the case of RM$_3$.}
\label{fig:pk_corner}
\end{figure}

With the radial model in hand, \texttt{PITSZI} can constrain the pressure fluctuation model parameters given the data. The power spectrum of the data is extracted from the map $S$ given by Equation~\ref{eq:signal_map_def}, or from the cross spectra between two independent maps. Data and model are then compared by processing the model following the formalism discussed in Section~\ref{sec:modeling}. 

Again, two fitting methods have been implemented: the MCMC parameter space sampling, and a non-linear least-squares fit. For the MCMC, the log-likelihood is defined, as for the profile, under the Gaussian approximation\footnote{In practice, the probability distribution followed by the data is not strictly Gaussian, especially for the first bins if the number of modes that are averaged to extract the spectrum is low.}
\begin{equation}
{\rm ln} \ \mathcal{L} \propto \sum_{m,n} \left(D - M\right)_m \left(C^{-1}\right)_{m,n} \left(D-M\right)_n,
\end{equation}
where $D$ corresponds to the power spectrum of the data, $M$ is the model power spectrum, and $C$ is the covariance matrix. The model is given for both methods by the sum of the different contributions to the power spectrum
\begin{equation}
M \equiv M(k_{\rm 2D}) = \mathcal{P}_{\rm ICM}(k_{\rm 2D}) + A_{\rm noise} \mathcal{P}_{\rm noise}(k_{\rm 2D}) + A_{\rm CIB} \mathcal{P}_{\rm CIB}(k_{\rm 2D}).
\end{equation}
Here, the quantity $\mathcal{P}_{\rm ICM}(k_{\rm 2D})$ refers to the ICM pressure model, which depends on the (fixed) pressure profile and on the pressure fluctuation parameters. The nuisance parameters $A_{\rm noise}$ and $A_{\rm CIB}$ allow \texttt{PITSZI} to marginalize over the amplitude of the noise and the CIB, to account for possible differences between the predictions and these components in the real data, and to check for degeneracies with the pressure fluctuation model. The power spectrum covariance matrix is given by the sum of the contributions from the noise and the CIB. The contribution from the pressure fluctuation can be added to account for sample variance using a reference spectrum that matches the data. In practice, we first fit the model without accounting for sample variance to obtain a reference model, and then refit the data accounting for the sample variance using the pressure fluctuation model covariance given the best-fit model from the first step. We check that the results are stable after only 1 iteration, given the relatively high signal-to-noise ratio for MACS~J0717.5+3745. The sample variance is thus accounted for in the total error budget.

We apply this method to the NIKA data on MACS~J0717.5+3745. We fit for the pressure fluctuation normalization $\sigma_{\mathcal{P}}$, the injection scale $L_{\rm inj}$, and marginalize over the noise ($A_{\rm noise}$) and CIB ($A_{\rm CIB}$) normalization. Given the scales sampled by the data, it is not feasible to fit for the slope without large degeneracies with other parameters, and we therefore fix it to $\alpha = -11/3$ \citep[see also][]{Dupourque2023,Dupourque2024}. We use flat priors on parameters, as $\sigma_{\mathcal{P}} \in [0,3]$, $L_{\rm inj} \in [50, 5000]$, $A_{\rm noise} \in [0,5]$ and $A_{\rm CIB} \in [0,5]$. However, this only affects the results for the CIB normalization since it is poorly constrained, so that we marginalize over its uncertain amplitude within a large range. We use RM$_3$ as our baseline radial model, but since this choice is the main source of uncertainty, we also provide the results for the other models. In terms of methodology, we make the following choices as a reference: the auto-spectrum is used (note that when using the cross-spectrum between two data subsets, the noise amplitude is fixed to zero since the two sets are independent); no kSZ correction is performed; the point sources are subtracted but not masked ; the weight map ($\omega$) is defined as being proportional to the radial model Compton parameter map multiplied by the binary mask (2 arcmin radius region of interest, plus point sources if relevant) and smoothed with a 20 arcsec FWHM Gaussian kernel, giving more weight to the inner part of the cluster; the non-linear least square fit is used; the projection approach is used. Nevertheless, we provide the results when varying all of these choices. For the non linear least square fit, the best fit parameters and their covariance matrix is used to produce multivariate Gaussian distributions as the constraint in the parameter space, accounting for sharp limits on the parameters.

Numerical results of the fits are listed in Table~\ref{tab:inference_fluctuation_results}. In Figure~\ref{fig:pk}, we show the constraints in terms of the 2D power spectrum and the corresponding constraints in terms of the 3D power spectrum, for our reference analysis choice RM$_3$. Figure~\ref{fig:pk_corner} provides the posterior constraints on the parameter space, focusing only on the non-extreme models RM$_2$, RM$_3$ and RM$_4$ for clarity. As expected, the value of $\sigma_{\mathcal{P}}$ and $L_{\rm inj}$ tends to decrease with model complexity, from RM$_1$ to RM$_5$, although the value of $L_{\rm inj}$ is nearly unconstrained for RM$_5$, which explains its high central value. Excluding the two extremes (RM$_1$ and RM$_5$), the results are relatively stable with respect to the choice of the radial model. This can also be observed in the 3D power spectrum constraint and the parameter space constraint. 

The statistical uncertainty on the normalization $\sigma_{\mathcal{P}}$ is of the order of 10\%, while the injection scale is also well constrained, with similar precision. If the injection scale is poorly-constrained, for instance because the power spectrum peak falls beyond the scales accessible from the data (RM$_1$) or because the fluctuations have been absorbed in the radial model (RM$_5$), the uncertainties  increase significantly. The systematic uncertainty due to the radial model is much larger than the statistical uncertainty when including all radial models (about 50\%), but these uncertainties are comparable when excluding the two extreme radial models (see also Appendix~\ref{app:radial_model_implication} for an attempt to quantify the systematics induced by these models). We also show that the analysis choices have a minor impact on the results as they are all nearly compatible within the statistical uncertainties. The main changes are observed in the case of the kSZ template subtraction and the change of the weighting scheme, which effectively slightly changes the power spectrum extraction region under consideration. They both imply a slightly higher pressure fluctuation amplitude (by about $3 \sigma$). Given the relatively high signal-to-noise detection of the pressure fluctuations, the parameter posterior distribution is nearly Gaussian, so that the MCMC fit and the non-linear least squares method are fully consistent for $\sigma_{\mathcal{P}}$, $L_{\rm inj}$ and $A_{\rm noise}$. The CIB normalization is compatible with unity, but is poorly constrained. As its posterior distribution significantly deviates from that of a Gaussian, the CIB constraints given in Table~\ref{tab:inference_fluctuation_results} for the non-linear least squares and MCMC fits differ. The noise normalization is well constrained and is compatible with unity, indicating that it describes the data well.

\section{Implications for the nonthermal ICM physics}\label{sec:physics}
The constraints on the pressure fluctuation power spectrum give information on the nonthermal physic of the ICM, assuming the fluctuations are directly related to the turbulence. In this Section, we explore the physical consequences of the measurements obtained above, compare the results with other work, and discuss limitations and caveats of the method. 

In \texttt{PITSZI}, the physical interpretation is implemented in the class \texttt{Physics} (see Figure~\ref{fig:overview}), essentially as a list of functions that encode the equations described in detail below, and the results from the literature used here. As will be discussed further later, we stress that the results derived here are subject to several assumptions and are limited by systematic uncertainties, in particular due to the choice of the radial model.

\begin{table*}[h]
\caption{\footnotesize{Nonthermal ICM physics derived from our reference analysis. The reported values correspond to the median of the distribution, and the uncertainties account for both the statistical uncertainties in the power spectrum measurement and the scatter in the relation derived by \citep{Zhuravleva2023}. The uncertainties obtained when ignoring the scatter are given in parenthesis.}}
\begin{center}
\resizebox{0.9\textwidth}{!} {
\begin{tabular}{c|ccccc}
\hline
\hline
Model & $\mathcal{M}_{\rm 3D}$ & $\sigma_v$ & $\frac{P_{\rm kin}}{P_{\rm kin} + P_{\rm th}}$ & $b_{{\rm HSE}, \Delta=2500}$$^{\dagger}$ & $b_{{\rm HSE}, \Delta=500}$$^{\dagger}$ \\
& (--) & (km/s) & (--) & (--) & (--) \\ [1mm]
\hline
\hline
RM$_1$ & $0.82^{+0.16}_{-0.16} \left(^{+0.04}_{-0.06} \ {\rm stat}.\right)$ & $1618^{+334}_{-323} \left(^{+74}_{-115} \ {\rm stat}.\right)$  & $0.27^{+0.08}_{-0.08} \left(^{+0.02}_{-0.03} \ {\rm stat}.\right)$ & $0.33^{+0.07}_{-0.06} \left(^{+0.007}_{-0.010} \ {\rm stat}.\right)$ & $0.41^{+0.05}_{-0.05} \left(^{+0.005}_{-0.008} \ {\rm stat}.\right)$ \\ [1mm]
RM$_2$ & $0.62^{+0.11}_{-0.11} \left(^{+0.02}_{-0.02} \ {\rm stat}.\right)$ & $1221^{+220}_{-230} \left(^{+41}_{-45} \ {\rm stat}.\right)$   & $0.17^{0.05}_{0.05} \left(^{+0.01}_{-0.01} \ {\rm stat}.\right)$   & $0.30^{+0.06}_{-0.06} \left(^{+0.004}_{-0.004} \ {\rm stat}.\right)$ & $0.38^{+0.05}_{-0.05} \left(^{+0.003}_{-0.003} \ {\rm stat}.\right)$ \\ [1mm]
RM$_3$ & $0.59^{+0.20}_{-0.19} \left(^{+0.02}_{-0.03} \ {\rm stat}.\right)$ & $1176^{+404}_{-385} \left(^{+43}_{-52} \ {\rm stat}.\right)$   & $0.16^{+0.10}_{-0.08} \left(^{+0.01}_{-0.01} \ {\rm stat}.\right)$ & $0.29^{+0.07}_{-0.06} \left(^{+0.004}_{-0.005} \ {\rm stat}.\right)$ & $0.38^{+0.05}_{-0.05} \left(^{+0.003}_{-0.004} \ {\rm stat}.\right)$ \\ [1mm]
RM$_4$ & $0.64^{+0.21}_{-0.21} \left(^{+0.04}_{-0.05} \ {\rm stat}.\right)$ & $1261^{+421}_{-413} \left(^{+87}_{-90} \ {\rm stat}.\right)$   & $0.18^{+0.10}_{-0.09} \left(^{+0.02}_{-0.02} \ {\rm stat}.\right)$ & $0.30^{+0.06}_{-0.06} \left(^{+0.008}_{-0.009} \ {\rm stat}.\right)$ & $0.39^{+0.05}_{-0.05} \left(^{+0.006}_{-0.007} \ {\rm stat}.\right)$ \\ [1mm]
RM$_5$ & $0.28^{+0.17}_{-0.14} \left(^{+0.10}_{-0.11} \ {\rm stat}.\right)$ & $545^{+332}_{-273}  \left(^{+205}_{-226} \ {\rm stat}.\right)$ & $0.04^{+0.06}_{0.03} \left(^{+0.04}_{-0.03} \ {\rm stat}.\right)$  & $0.24^{+0.07}_{-0.06} \left(^{+0.021}_{-0.023} \ {\rm stat}.\right)$ & $0.34^{+0.06}_{-0.05} \left(^{+0.017}_{-0.021} \ {\rm stat}.\right)$ \\ [1mm]
\hline
\end{tabular}
}
\end{center}
{\footnotesize {\bf Notes.} 
$^{\dagger}$ These results were obtained with the scaling of Equation~\ref{eq:Z23_scaling_1} (sum) but no significant change is observed when using Equation~\ref{eq:Z23_scaling_2} (product) instead.
}
\label{tab:inference_nt_results}
\end{table*}

\begin{figure*}
        \centering
        \includegraphics[width=0.99\textwidth]{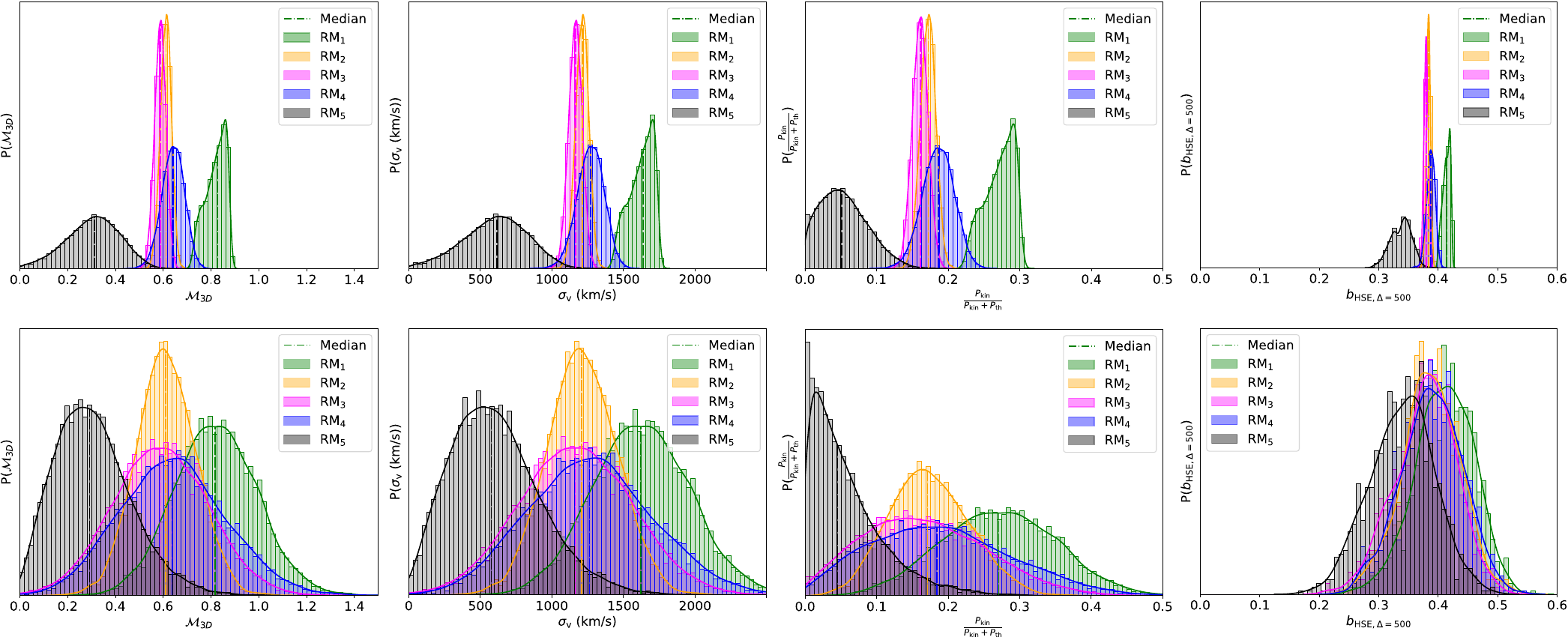}
        \caption{Constraints on the nonthermal ICM parameters. From left to right: constraints on the 3D Mach number, the velocity dispersion, the kinetic to total pressure, and the hydrostatic mass bias at $R_{500}$ . The constraints for the different radial models are shown with different colors, as indicated in the legend. The top panels provide the constraints obtained without accounting for the scatter in the relation by \citep{Zhuravleva2023}, while in the bottom panels this scatter is accounter for.}
\label{fig:pdf_nt_param}
\end{figure*}

\subsection{From the power spectrum to nonthermal physics}
%
The power spectrum of the pressure fluctuations is related to the turbulent 3D Mach number, as characterized by numerical simulations \citep[e.g.,][]{Gaspari2014,Mohapatra2020,Mohapatra2021,Mohapatra2022,Zhuravleva2023}. For instance, assuming adiabatic perturbations, the relation obtained by \cite{Gaspari2014} for density perturbations becomes, when expressed as pressure perturbations
\begin{equation}
    \mathcal{M}_{\rm 3D} = \frac{4}{\gamma} A_{\delta P / \bar{P}}(k_{\rm peak}) \left(\frac{L_{\rm inj}}{500 \ {\rm kpc}}\right)^{-s},
    \label{eq:G14_scaling}
\end{equation}
where $A_{\delta P / \bar{P}}(k_{\rm peak})$ is the peak of the characteristic amplitude of the pressure fluctuation power spectrum (Equation~\ref{eq:charac_amplitude}). The slope $s \simeq 0.25$ gives a small correction as a function of injection scale. In contrast, \cite{Zhuravleva2023} provide a scaling of the Mach number with the integrated power spectrum ($\propto \sigma_{\mathcal{P}}$), for pressure perturbations, accounting for a radial dependence and for different dynamical states. These relations read
\begin{equation}
    \frac{\mathcal{M}_{\rm 3D}}{\sqrt{3}} = a + b \times \frac{r}{R_{500}} + c \times \frac{\delta \xi}{\xi}, 
    \label{eq:Z23_scaling_1}
\end{equation}
or 
\begin{equation}
    \frac{\mathcal{M}_{\rm 3D}}{\sqrt{3}} = a \times \left(\frac{r}{R_{500}}\right)^{b} \times  \left(\frac{\delta \xi}{\xi}\right)^{c}, 
    \label{eq:Z23_scaling_2}
\end{equation}
where $a$, $b$ and $c$ are coefficients that depend on the dynamical state and the choice of scaling relation (see their Table 1 and their Fig.~6, for the pressure). Here, we have converted our normalization to the natural logarithm based normalization used by \citet{Zhuravleva2023}, $\sigma_{\rm ln} = \sqrt{{\rm log}\left(\left(1+\sqrt{1 + 4 \sigma_{\mathcal{P}}^2}\right)/2\right)}$, and we have used their notation, $\delta \xi / \xi = 2 \sqrt{2 {\rm log}\left(2\right)} / {\rm log}\left(10\right) \sigma_{\rm ln}$. Applying the results from \cite{Zhuravleva2023}, ignoring the radial dependence and accounting for the ellipticity of our models (their Fig.~6), we obtain the 3D Mach number values listed in Table~\ref{tab:inference_nt_results};  Figure~\ref{fig:pdf_nt_param} shows the corresponding probability distribution of the Mach number. For the intermediate model RM$_3$, $\mathcal{M}_{\rm 3D} \sim 0.6$. Error bars were computed by propagating the uncertainty on the power spectrum parameters and including the scatter from the relation by \cite{Zhuravleva2023} via Monte Carlo realizations. As we can see, the scatter in the relation dominates over the statistical error, but it remains smaller than the systematic effect due to the choice of the radial model.

The 3D Mach number is related to the velocity dispersion $\sigma_v$ via
\begin{equation}
	\mathcal{M}_{\rm 3D} = \frac{\sigma_v}{c_{\rm s}},
\end{equation}
where $c_{\rm s}$ is the sound speed, which depends on the gas temperature $T_{\rm gas}$ as
\begin{equation}
	c_{\rm s} = \sqrt{\frac{\gamma k_{\rm B}  T_{\rm gas}}{\mu_{\rm gas} m_{\rm p}}}.
\end{equation}
Here $\gamma = 5/3$ is the gas adiabatic index, $k_{\rm B}$ the Boltzmann constant, $\mu_{\rm gas} \simeq 0.61$ the gas mean molecular weight and $m_{\rm p}$ the mass of the proton. Here we assume a mean temperature $T_{\rm gas} = 15$ keV \citep{Adam2017b}. This implies that $c_{\rm s} = 1981$ km/s. We list the corresponding values in Table~\ref{tab:inference_nt_results} and show the probability distribution of this parameter in Figure~\ref{fig:pdf_nt_param}. For the intermediate model RM$_3$, we obtain $\sigma_v \sim 1200$ km/s.

The ratio between kinetic and thermal ICM pressure is related to the Mach number as
\begin{equation}
    X = \frac{P_{\rm kin}}{P_{\rm th}} = \frac{1}{2} \gamma \left(\gamma-1\right) \mathcal{M}_{\rm 3D}^2.
    \label{eq:kin_to_tot_pressure}
\end{equation}
We use this equation to derive the kinetic to kinetic plus thermal pressure ratio. Our findings are listed in Table~\ref{tab:inference_nt_results}, and are displayed in Fig.~\ref{fig:pdf_nt_param}. We obtain $P_{\rm kin} / P_{\rm th+kin} \sim 0.16$ for RM$_3$.

The hydrostatic equilibrium assumption is often used to obtain estimates of the mass of galaxy cluster via \citep{Pratt2019}
\begin{equation}
    M_{\rm HSE}(<r) = -\frac{r^2}{G \rho_{\rm gas}(r)} \frac{d P_{\rm th}(r)}{dr},
\end{equation}
using the gas density, $\rho_{\rm gas}$, and thermal pressure estimates, $P_{\rm th}$, derived from SZ or X-ray observations, with $G$ being Newton's constant.
However, turbulent gas motions will contribute to balance the gravitational field and will induce a bias in the hydrostatic mass if they are non-negligible. Turbulent motions expected to dominate the hydrostatic mass bias \citep{Pratt2019}. Thanks to Equation~\ref{eq:kin_to_tot_pressure}, we can compute a corrected mass profile, $M_{\rm corr}$, by substituting the thermal pressure with the thermal plus kinetic pressure: $P_{\rm th}(r) \rightarrow P_{\rm th}(r) + P_{\rm kin}(r) = \left(1+X(r)\right) P_{\rm th}(r)$. The mass bias profile is then obtained from
\begin{equation}
    b_{\rm HSE} = 1 - \frac{M_{\rm HSE}(<r)}{M_{\rm corr}(<r)}.
\end{equation}
The mass bias at an overdensity $\Delta$ is given by $b_{{\rm HSE}, \Delta} = 1 - \frac{M_{\rm HSE}(R_{\Delta, {\rm HSE}})}{M_{\rm corr}(R_{\Delta, {\rm corr}})}$, where $R_{\Delta}$ is estimated by integrating the mass profile such that the enclosed density within $R_{\Delta}$ reaches $\Delta$ times the critical density of the Universe. In the following, we perform our calculations by using as a reference the gas density profile obtained from XMM-Newton observation used in \cite{Adam2017b} and the pressure profile of model RM$_1$, as it is centered on the X-ray peak. The kinetic pressure correction, $X$, is obtained via Equation~\ref{eq:kin_to_tot_pressure} using the Mach number derived from the scaling by \cite{Zhuravleva2023} given in Equation~\ref{eq:Z23_scaling_1}, as a function of radius. The statistical uncertainties arising from the pressure fluctuation power spectrum are also accounted for, but these are subdominant. Our estimates of the kinetic pressure correction, except for model RM$_1$, use radial pressure models for which the center is not fully consistent with that for the thermal model (density and pressure). However, the results obtained for the hydrostatic mass bias are barely affected by the exact choice of thermal model. This is because to first order the bias corresponds to a relative difference with respect to the perfect hydrostatic equilibrium model, such that changes in the exact reference model are not important, at least compared to other uncertainties. Our findings are listed in Table~\ref{tab:inference_nt_results} and shown in Figure~\ref{fig:pdf_nt_param}. We obtain $b_{{\rm HSE}, \Delta=2500} \sim 0.3$ and $b_{{\rm HSE}, \Delta=500} \sim 0.4$.

\subsection{Discussions: comparison with previous work}\label{sec:Discussions_comparison_previous_works}
Recently, \cite{Heinrich2024} used \textit{Chandra} data to infer the ICM turbulence from X-ray surface brightness density fluctuations. In their sample of 80 clusters, they found the highest characteristic velocities for MACS~J0717.5+3745. They obtained a Mach number $\mathcal{M}_{\rm 3D} \simeq 0.47$, a velocity dispersion in the range $\sim$ 1000 -- 1300 km/s, and a kinetic to total pressure fraction $P_{\rm kin}/\left(P_{\rm kin} + P_{\rm th}\right)$ in the range $\sim$ 7.8--18.9\%. These results are in excellent qualitative agreement with our findings (Table~\ref{tab:inference_nt_results}), indicating that we are indeed probing the same physical process with these SZ observations, and that systematic effects are limited. In fact, combining the SZ pressure and X-ray density fluctuations, it is possible to constrain the thermodynamic nature of the turbulence itself. The pressure ($P$) and density ($\rho$) perturbations are related via
\begin{equation}
\frac{\delta P}{\bar{P}} = \Gamma \frac{\delta \rho}{\bar{\rho}},
\end{equation}
with $\Gamma = [0, 1, \frac{5}{3}]$ for isobaric, isothermal, and adiabatic fluctuations, respectively. \cite{Heinrich2024} measured $\left(\delta \rho / \bar{\rho} \right)_k \equiv A_{\delta \rho / \bar{\rho}}(k_{\rm 3D}) = \sqrt{4 \pi k_{\rm 3D}^3 \mathcal{P}_{\rho, {\rm 3D}}(k_{\rm 3D})}$ at $R_{2500} = 507$ kpc for MACS~J0717.5+3745, obtaining $A_{\delta \rho / \bar{\rho}}(1/R_{2500}) = 0.329^{+0.164}_{-0.110}$. Combined with our measurement at the same scale, $A_{\delta P / \bar{P}}(1/R_{2500}) = [0.47^{+0.04}_{-0.02}, 0.39^{+0.02}_{-0.02}, 0.37^{+0.02}_{-0.02}, 0.40^{+0.03}_{-0.04}, 0.15^{+0.06}_{-0.06}]$ we estimate $\Gamma = [1.43^{+0.72}_{-0.50}, 1.18^{+0.59}_{-0.40}, 1.12^{+0.56}_{-0.38}, 1.22^{+0.62}_{-0.42}, 0.46^{+0.29}_{-0.25}]$ for models RM$_1$ to RM$_5$, respectively. These values agree with those expected for adiabatic or isothermal perturbations, but exclude isobaric perturbations (except for RM$_5$, which sits between isobaric and isothermal). Nevertheless, we note that the two analyses are not necessarily self-consistent since the radial model may be incompatible (e.g. the cluster center differs). In the future, a joint analysis of SZ and X-ray perturbations at the surface brightness level, with joint radial modeling, may be effective in better constraining the nature of the ICM fluctuations.

We also compared the velocity dispersion obtained via the pressure fluctuations measurement to that obtained directly using the kinetic SZ effect \citep{Sayers2013,Adam2017a}. We used the ICM line-of-sight velocity maps derived by \cite{Adam2017a} under the assumptions of their two ICM density models F1 and F2. The image standard deviation, extracted within their mask, which roughly matches our region of interest, gives $\sigma_{\rm kSZ} \simeq 1500$ km/s and $\sigma_{\rm kSZ} \simeq 1300$ for their models F1 and F2, respectively. Although the comparison is limited because of projection effects, noise contamination in the kSZ velocity image, and because the kSZ is only sensitive to the line-of-sight velocity component, the two measurements compare well. This could indicate that the velocity measured via kSZ, which is essentially dominated by the bulk velocity of the sub-cluster components, corresponds well to the injection of turbulence on the largest scales, that will eventually  cascade and dissipate at smaller scales.

\begin{figure*}
        \centering
        \includegraphics[width=0.49\textwidth]{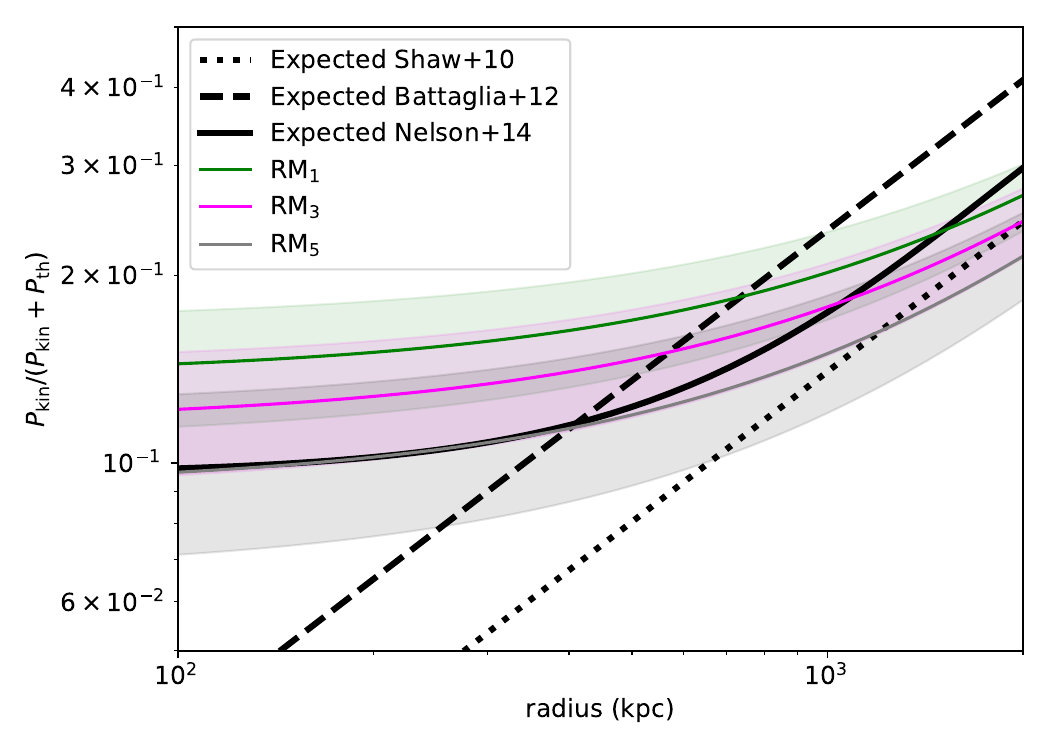}
        \includegraphics[width=0.49\textwidth]{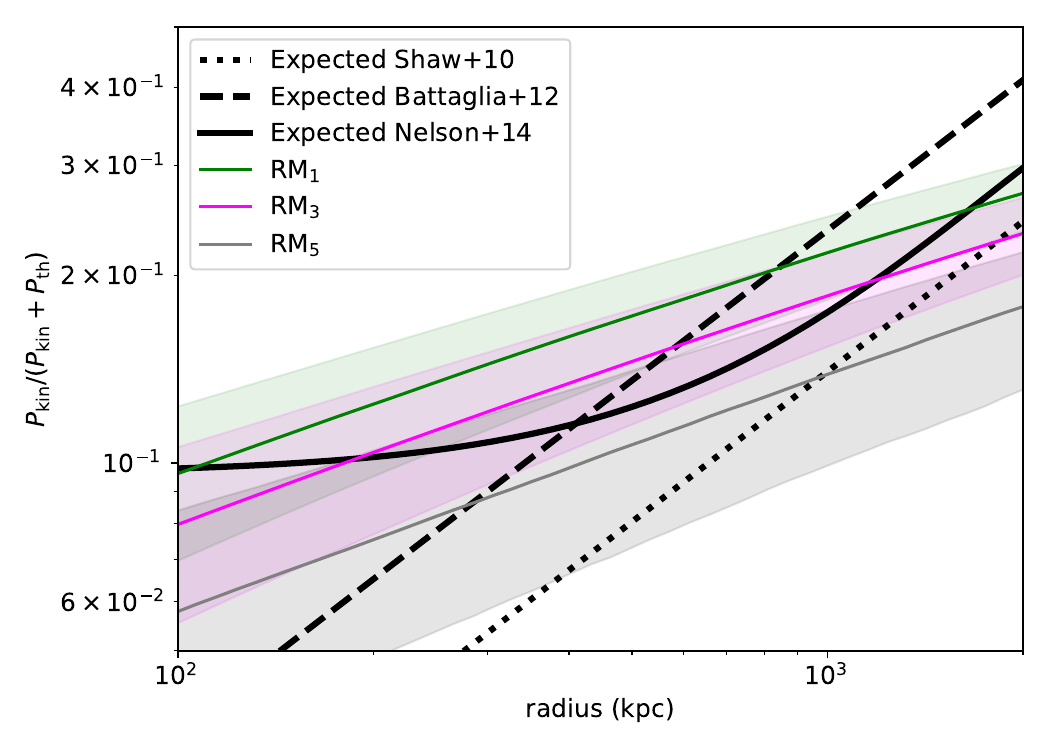}
        \includegraphics[width=0.49\textwidth]{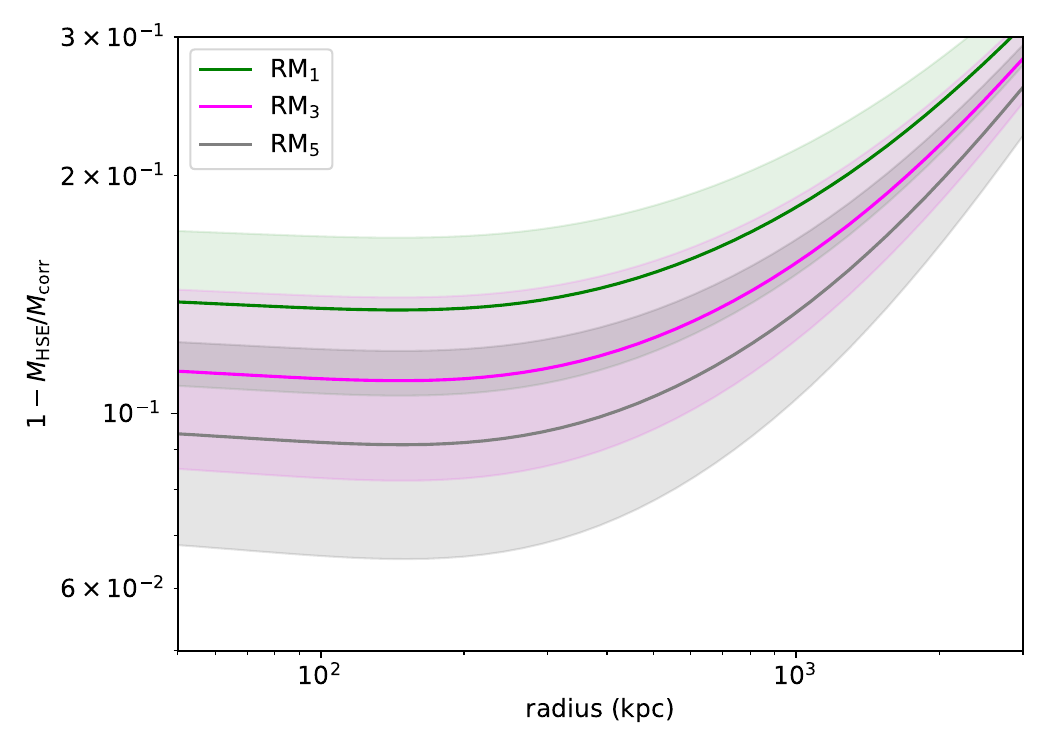}
        \includegraphics[width=0.49\textwidth]{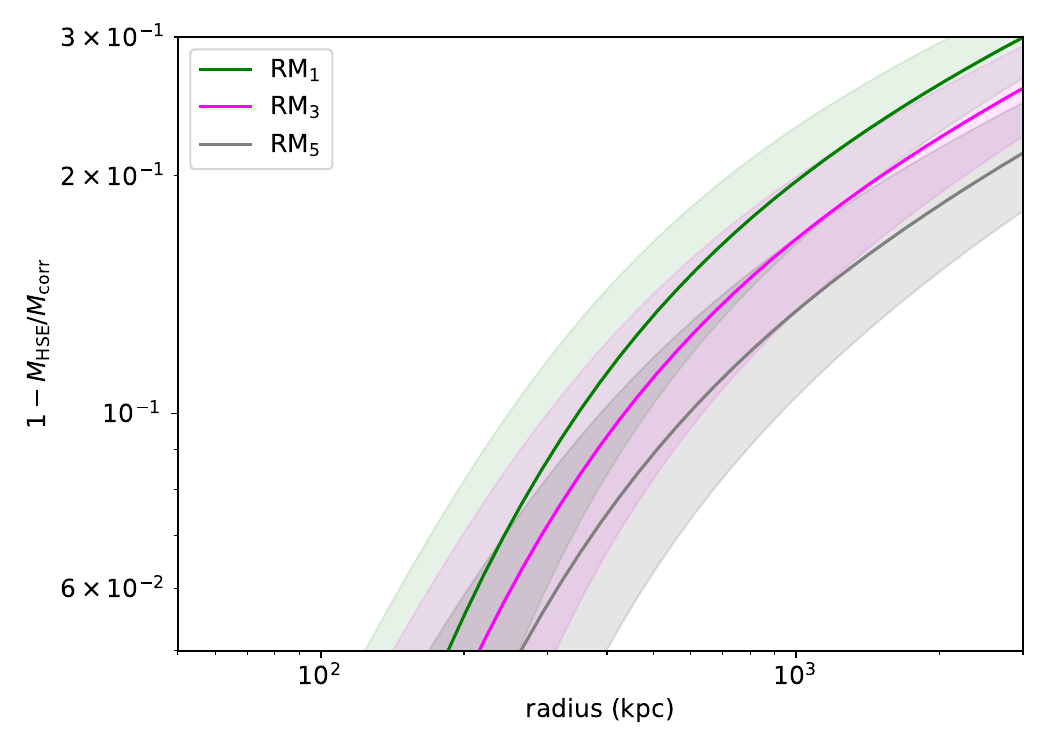}
        \caption{Kinetic to thermal plus kinetic pressure ratio and hydrostatic mass bias as a function of radius. Constraints are obtained using the scaling relation of \cite{Zhuravleva2023} given in Equation~\ref{eq:Z23_scaling_1} (sum, left panel) and Equation~\ref{eq:Z23_scaling_2} (product, right panel). Error contours account for the statistical uncertainties in the power spectrum measures and the scatter in the relation by \cite{Zhuravleva2023}.
        {\bf Top}: Kinetic to thermal plus kinetic pressure ratio constraint and comparison to expectations from numerical simulations \citep{Shaw2010,Battaglia2012,Nelson2014}. Here we assumed $M_{500} = 26.5 \times 10^{14}$ M$_{\odot}$ according to the result from model RM$_3$, but this choice does not significantly affect our results.
        {\bf Bottom}: Hydrostatic mass bias profile computed as $b \equiv 1 - \frac{M_{\rm HSE}}{M_{\rm corr}}$.
        The results from RM$_2$ and RM$_4$ were omitted for clarity, but they lie between models RM$_1$ and RM$_3$, and RM$_3$ and RM$_5$, respectively.
        }
\label{fig:physics_pressure_fraction}
\end{figure*}
The kinetic to thermal pressure support in galaxy clusters has been predicted using numerical simulations \citep{Shaw2010,Battaglia2012,Nelson2014}. Using the relation between Mach number and the pressure fluctuation power spectrum obtained by \cite{Zhuravleva2023}, we can estimate this quantity via Equation~\ref{eq:kin_to_tot_pressure} and compare our results with expectations. The comparison is shown in Figure~\ref{fig:physics_pressure_fraction}, where we account for both statistical uncertainties in the power spectrum constraint and the scatter in the relation by \cite{Zhuravleva2023} that we use. As we can see, our estimates compare well with expectations, although we are limited by the uncertainty in the shape of the radial dependence both in the predictions from numerical simulations and in the scaling relation that we use. Apart from model RM$_5$, our results are generally slightly above the predictions, which might be anticipated since MACS~J0717.5+3745 is one of the most dynamically active clusters in the Universe. 

In Figure~\ref{fig:physics_pressure_fraction}, we also derived a constraint on the hydrostatic mass bias profile, for which we observed a strong radial shape dependence on the choice of the scaling relation used to obtain the constraint. The two converge at $r \gtrsim 300$ kpc and we obtain a bias of $\sim$ 0.15--0.3. When comparing directly the recovered masses at overdensity $\Delta$, this increases to $b_{{\rm HSE}, \Delta=2500} \sim 0.3$ and $b_{{\rm HSE}, \Delta=500} \sim 0.4$. Such a value corresponds to that generally assumed to reconcile CMB cosmological constraints with cluster counts from Planck \citep[e.g.,][]{PlanckXX2014}. However, it is obtained for a single cluster exhibiting  extreme dynamical activity, that is not representative of the full population. Application of the same analysis to other clusters is likely to lead to much lower values for the bias. Moreover, these results are affected by strong assumptions and limitations, which we discuss next, and which prevent us from drawing any strong conclusions.

\subsection{Discussion: limitations, approximations and systematic effects}
The results presented in this paper are intrinsically limited by choices in the analysis and systematic effects, in addition to assumptions regarding the nature of the signal that we are actually probing. It should again be stressed that MACS~J0717.5+3745 is an extremely complex system. While it is an excellent target to develop the methodology presented in this paper, the signal that we are extracting is likely to be affected not only by turbulent motions, but also by bulk motions, clumping, or a residual kSZ signal. 

Considering MACS~J0717.5+3745 as a single cluster is also a strong approximation: should it be thought as a single object with strong internal activity and strong pressure fluctuations, as we did, or should it be considered as a set of still nearly independent sub-clusters? Consequently, what part of the signal should be included in the hydrostatic component and what part should go to the fluctuations? This question reflects in the issue of the choice of the radial model, which has been intensively discussed in the literature. It is an obvious issue for MACS~J0717.5+3745, but it should affect all clusters to some extent, not only strong mergers \citep[e.g.,][for discussion]{Romero2023}. Here we have attempted to estimate the associated uncertainty by applying a number of radial models and bracketing our results with two extreme models. Despite these limitations, our results provide a step forward in the attempt of characterizing the nonthermal physics of the ICM from resolved SZ data, and we can still compare our finding with alternative observations.

In the present paper, we have implemented an analysis framework based on the projection of the 3D pressure fluctuation power spectrum to the 2D SZ fluctuation power spectrum. This involves a linear scaling between the 2D and the 3D spectra, which is completely accurate only for a flat surface brightness and at scales that are sufficiently small, particularly compared to the injection scale. While we have shown that this method leads to only $\lesssim 10\%$ bias on the model predictions in our context, future work will likely benefit from avoiding such approximations to obtain a more reliable model in any configuration and at all scales. This could be done via the use of simulation based inference techniques, as applied in \cite[e.g.][]{Dupourque2023,Dupourque2024}.

The pressure fluctuation power spectrum measure is directly obtained from SZ data. When deriving the nonthermal ICM physical properties in Section~\ref{sec:physics}, we relied on scaling relations, calibrated from numerical simulations,  to connect the pressure fluctuation power spectrum to the turbulence via the 3D Mach number. Here we have focused on the relations calibrated by \cite{Zhuravleva2023}, but testing the implications of different relations based on different simulations could be valuable. We also noted that the scatter in the relation between the pressure fluctuation power spectrum and the 3D Mach number was one of the main sources of uncertainty when extracting nonthermal information on the ICM.

In the present work, we did not attempt to measure the evolution of the power spectrum in different radial bins because we were already limited in terms of the spatial scales accessible with the NIKA data. This implied that our measurement was only used to fix the normalization of the 3D Mach number, and we relied on the radial trend given by the scaling relations from \cite{Zhuravleva2023} when deriving the nonthermal radial profiles. In the future, the NIKA2 data should allow us to probe larger scales, owing to the larger field of view of its camera \citep{Adam2018a}, thus allowing us measure pressure fluctuations further out of the cluster core and split the data into different radial bins to directly constrain the radial evolution from the data.

We used a fixed mean gas temperature across the cluster when extracting the turbulent velocity dispersion. However, MACS~J0717.5+3745 contains complex temperature structure which significantly varies over the core region \citep{Adam2017b}. This implies that even when using a fixed Mach number, the velocity dispersion should present significant spatial variations, which we did not consider here.

Finally, and as already mentioned in Section~\ref{sec:Discussions_comparison_previous_works}, the combination of X-ray and SZ fluctuation measures is in principle very valuable to constrain the nature of the fluctuations. However, we only used the measured constraints on the power spectra and combined pressure and density fluctuation estimates a posteriori. In addition to the coherence of the radial model that is not guaranteed here, such a combination does not allow one to truly test that the measured fluctuations arise from the same physical origin. In the future, applying cross correlation techniques between the different observables should be efficient in testing the true nature of the fluctuations, and in limiting systematic effects that are not correlated between SZ and X-ray observables.

\section{Summary and conclusions}\label{sec:Summary_and_conclusions}
In this paper, we presented new developments made to measure the pressure fluctuation power spectrum of the diffuse gas in galaxy clusters using resolved SZ data. This led to the construction of the \texttt{PITSZI} software suite, which is publicly available at \url{https://github.com/remi-adam/pitszi}. \texttt{PITSZI} allows one to model the ICM pressure distribution, accounting for the radial (possibly triaxial) component and the fluctuations (gaussian or lognormal) described by a power spectrum. The software can be used to produce mock data and to constrain SZ  pressure fluctuations via different fitting methods. \texttt{PITSZI} has been developed and extensively tested with NIKA and NIKA2 data, but it can in principle work with any SZ imaging data (e.g., Planck, SPT, ACT, Bolocam, MUSTANG, MUSTANG2, NIKA, NIKA2). 

The code was applied to the triple merging cluster MACS~J0717.5+3745. We summarize our main findings below.
\begin{itemize}
\item The methodological framework developed in \texttt{PITSZI} improves over previous analyses. In particular, the power spectrum model prediction was shown to be accurate to within less than 10\% for realistic ICM models. Additionally, \texttt{PITSZI} allows us directly to constrain the 3D pressure fluctuation power spectrum parameters using a Bayesian approach.
\item We reported a high significance detection of pressure fluctuations within $\sim \theta_{500}/2$. By modeling and fitting the power spectrum with an exponential cutoff powerlaw with a single injection scale, we obtained a large amplitude, $\sigma_{\mathcal{P}} \simeq 0.6$. The injection scale, $L_{\rm inj} \simeq 800$ kpc corresponds to expectations from merger driven energy injection. Given the scales sampled by the data and the sensitivity that was achieved, it was not possible to constrain the slope, which was fixed to the canonical value of $\alpha = -11/3$.
\item Because of projection effects, the instrument response and the noise in the data, it was not possible to distinguish Gaussian from lognormal perturbations despite the large amplitude of the fluctuations. As projection effects dominate, we anticipate that it will be extremely challenging to differentiate the two, even with improved observations. This result contrasts with the measurement reported in \cite{Khatri2016} for the Coma cluster with Planck.
\item The origin and nature of the detected fluctuations may be questionable in such a complex merging system, particularly given the presence of a known kSZ signal, and the fact that the choice of the radial (smooth) model plays a crucial role regarding this issue. We have provided a first attempt to quantify the induced systematic effects. We found that fitting a radial model to the data filters the fluctuations by a factor that increases with increasing model complexity, reaching 80\% for the most complex test models. On the other hand, a model that is too simplistic was found to induce spurious fluctuations that could overshoot the true signal, even for large fluctuations for the most simple model. These systematic effect were found to act mainly on scales $k \lesssim 1/L_{\rm inj}$. By testing radial models with different complexity, we have attempted to bracket the  systematics linked to the choice of radial model, but we conclude that this issue is still the most critical when performing quantitative fluctuation analyses.
\item Assuming that the measured pressure fluctuations are related to turbulence, the estimated power spectrum of MACS~J0717.5+3745 implies a 3D Mach number of about $\mathcal{M}_{\rm 3D} \sim 0.6$. This corresponds to a kinetic to thermal pressure fraction $P_{\rm kin} / P_{\rm th+kin} \sim 0.2$. Given the cluster mean temperature, this implies a very large velocity dispersion ${\sigma}_{v} \sim 1200$ km/s. Our   measurement aligns very well with previous constraints from X-ray surface brightness fluctuations based on the ICM density. This may indicate that the ICM perturbations are likely adiabatic in nature. Interestingly, we note that the turbulent velocity dispersion constraint also qualitatively matches the direct measurement obtained from kinetic SZ measures \citep{Sayers2013,Adam2017a}.
\item Assuming that turbulence is the main driver of the hydrostatic mass bias, we derive a value of $b_{\rm HSE} \sim 0.3-0.4$, depending on radius and on the considered model.
\end{itemize}

Future high resolution X-ray spectroscopic data will allow us to calibrate the relation between the pressure fluctuation spectrum and the true turbulent velocity spectrum using real data. As high resolution X-ray spectroscopy will remain much more expensive than deep SZ imaging, especially at high redshift, pressure fluctuation analyses are likely to become an important tool for investigating ICM turbulence. In fact, the methodology presented in this paper should be applied to any other cluster sample. \texttt{PITSZI} is in continuous development. In particular, the implementation of simulation based inference methods \citep{Dupourque2023,Dupourque2024} are ongoing, with the aim of accounting for sample variance in a more efficient way, which is crucial when dealing with cluster samples. In addition, joint constraints from SZ plus X-ray images are also being considered, and could provide further interesting insights into the nature of the perturbations at play.


\begin{acknowledgements}
This work is based on observations carried out under project number 237-13 and 222-14 with the NIKA camera at the IRAM 30 m Telescope. IRAM is supported by INSU/CNRS (France), MPG (Germany) and IGN (Spain).
The data are publicly available at \url{https://lpsc.in2p3.fr/NIKA2LPSZ/nika2sz.release.php}. See \cite{Adam2018b} for the corresponding article.


This work was supported by the French government through the France 2030 investment plan managed by the National Research Agency (ANR), as part of the Initiative of Excellence of Université Côte d'Azur under reference number ANR-15-IDEX-01. 
This work has been supported by the French government, through the UCA$^{\rm J.E.D.I.}$ Investments in the Future project managed by the National Research Agency (ANR) with the reference number ANR-15-IDEX-01.
This work was supported by the Programme National Cosmology et Galaxies (PNCG) of CNRS/INSU with INP and IN2P3, co-funded by CEA and CNES.
EP and NC acknowledge support from the French Agence Nationale de la Recherche (ANR), under grant ANR-22-CE31-0010. GWP acknowledges support from CNES, the French space agency.

We would like to thank the IRAM staff for their support during the campaigns. 
The NIKA dilution cryostat has been designed and built at the Institut N\'eel. In particular, we acknowledge the crucial contribution of the Cryogenics Group, and  in particular Gregory Garde, Henri Rodenas, Jean Paul Leggeri, Philippe Camus. 
This work has been partially funded by the Foundation Nanoscience Grenoble, the LabEx FOCUS ANR-11-LABX-0013 and the ANR under the contracts 'MKIDS' and 'NIKA'. 
This work has benefited from the support of the European Research Council Advanced Grants ORISTARS and M2C under the European Union's Seventh Framework Programme (Grant Agreement nos. 291294 and 340519).

We thank Instituto de Astrofisica de Andalucia (IAA-CSIC), Centro de Supercomputacion de Galicia (CESGA) and the Spanish academic and research network (RedIRIS) in Spain for hosting Uchuu DR1, DR2 and DR3 in the Skies \& Universes site for cosmological simulations. The Uchuu simulations were carried out on Aterui II supercomputer at Center for Computational Astrophysics, CfCA, of National Astronomical Observatory of Japan, and the K computer at the RIKEN Advanced Institute for Computational Science. The Uchuu Data Releases efforts have made use of the skun\@IAA\_RedIRIS and skun6\@IAA computer facilities managed by the IAA-CSIC in Spain (MICINN EU-Feder grant EQC2018-004366-P).

This research made use of Astropy, a community-developed core Python package for Astronomy \citep{Astropy2013}, in addition to NumPy \citep{VanDerWalt2011}, SciPy \citep{Jones2001}, and Ipython \citep{Perez2007}. Figures were generated using Matplotlib \citep{Hunter2007}. 
\end{acknowledgements}

\bibliographystyle{aa}
\bibliography{biblio}

\begin{appendix}
\section{Contribution from diffuse backgrounds and foregrounds}\label{app:unresolved_background}
\subsection{Cosmic Infrared Background (CIB)}\label{app:unresolved_background_cib}
\begin{figure}
        \centering
        \includegraphics[width=0.49\textwidth]{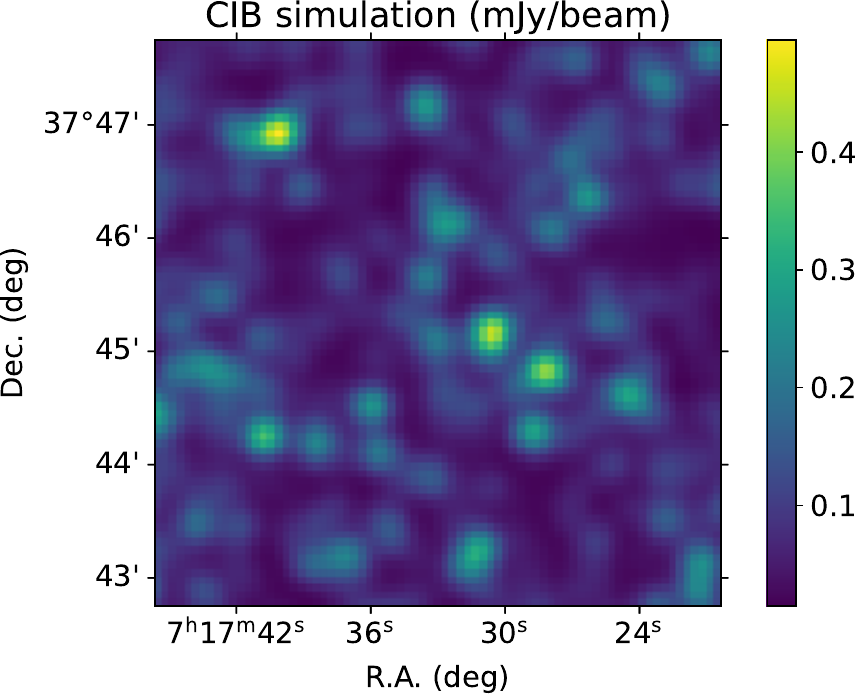}
        \caption{Example of a CIB realization at 150 GHz obtained using the \texttt{PYSIDES} code \citep{Bethermin2022}.}
\label{fig:cib_simulation}
\end{figure}
The CIB is one of the major contaminants in SZ fluctuation analyses, at least in the case of the scales probed with NIKA data. We use the method developed in \cite{Bethermin2017} to construct the mock realizations used in this paper. We apply the \texttt{PYSIDES} code \citep{Bethermin2022} to the UCHUU simulations \citep{Ishiyama2021}. We obtain a source catalog with sky positions, redshift and the NIKA 150 GHz fluxes over a 2 deg$^2$ region. We then split the catalog into 1000 independent regions of $10 \times 10$ arcmin$^2$. From each of them, a CIB map was generated by summing the contribution from the sources in the catalog, which we modeled given their flux, relative position with respect to the center, and the NIKA 150 GHz Gaussian beam. Catalog entries for which the flux is above the NIKA detection threshold \citep[taken as 0.4 mJy, corresponding to $2\sigma$,][]{Adam2017a} were removed to avoid including sources that would be subtracted or masked in real data. In Figure~\ref{fig:cib_simulation}, we show an example of one of the CIB realizations. These Monte Carlo realizations were used to model and account for the CIB along the analysis.

\subsection{Undetected low mass halos}\label{app:unresolved_background_halos}

We also considered the contribution from undetected low mass galaxy clusters in the field. We estimated the expected number of clusters in a $30 \times 30$ arcmin$^2$ field by integrating the \cite{Tinker2008} mass function across bins of redshift ($0 < z < 3$) and mass ($2\times10^{13}{\rm M}_{\odot} < {\rm M}_{500} < 3\times10^{15}{\rm M}_{\odot}$). We then drew the number of clusters from a Poisson distribution based on this estimate, and assigned each cluster a mass and redshift drawn from the mass function, along with a random position in the field. We repeated this process to obtain a 1000 mock cluster catalogs. Assuming spherical symmetry, we modeled the cluster's electronic pressure profile with a gNFW profile \citep{Nagai2007}, which was then integrated along the line of sight to obtain the Compton parameter profile of the cluster. By summing each cluster's contribution we obtained simulated maps of the tSZ signal. The modeling was undertaken using the \texttt{minot} Python package \citep{Adam2020}. The maps were convolved with the NIKA instrument response function (beam and transfer function) and used to compute its associated power spectrum as described in Section~\ref{sec:noise}. We found that given the present analysis framework and the scales that we probe, the contribution from undetected low mass clusters is about two orders of magnitude lower than that of the CIB, and therefore fully negligible.

\section{Uncertainties and systematics in the recovered power spectrum due to the methodology}\label{app:precision_of_the_model}
\begin{figure}
        \centering
        \includegraphics[width=0.49\textwidth]{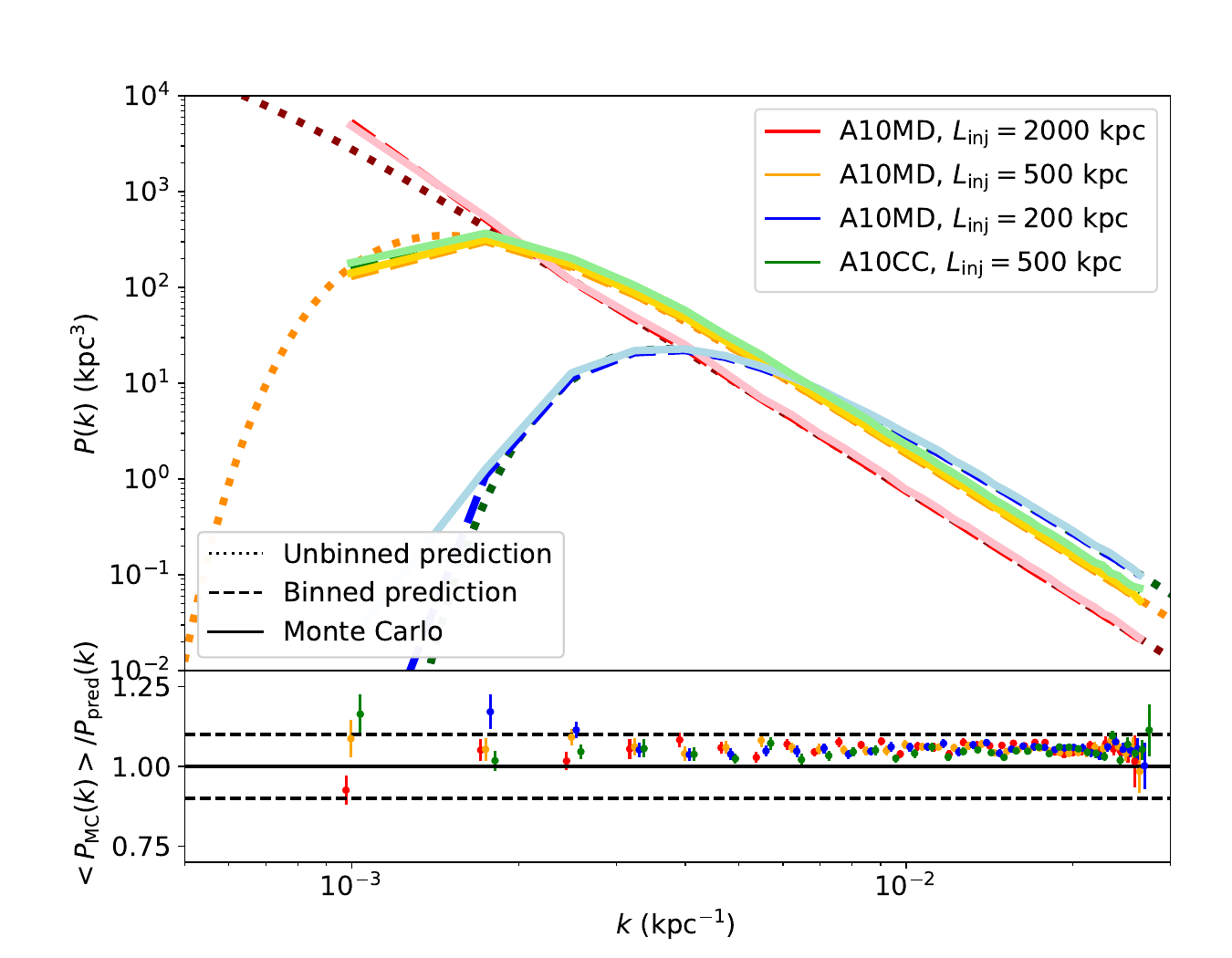}
        \includegraphics[width=0.49\textwidth]{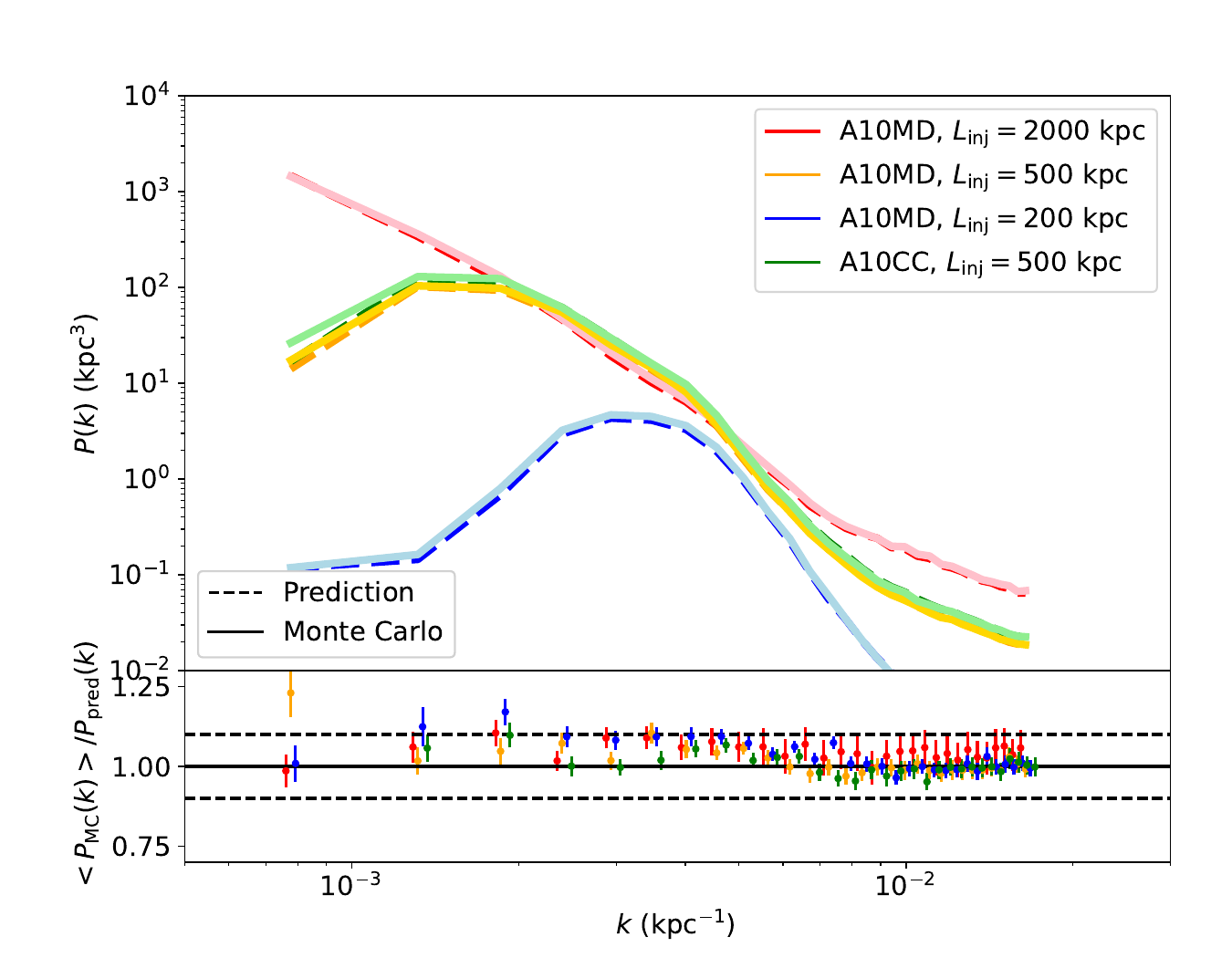}
        \caption{
        Comparison between the expected and the recovered (mean of Monte Carlo realizations) pressure fluctuation power spectra. The bias, shown at the bottom of each panel, is defined as the ratio between the two and we show the $\pm 10$\% interval as a dashed line. Error bars correspond to the uncertainties on the mean associated with the variability of Monte Carlo realizations.
        {\bf Top}: comparison after projection only, i.e. we are testing the validity of Equation~\ref{eq:projection_only_modeling}. The bias is larger than a factor of 1.3 only for the lowest bin for the blue curve, at scales well above the peak.
        {\bf Bottom:} comparison when including projection effects, the instrument response function and masking, i.e. we are testing the framework discussed in Section~\ref{sec:modeling}.
        The labels A10MD and A10CC refer to the morphologically disturbed and cool-core pressure profile models of \cite{Arnaud2010}, respectively.
        }
\label{fig:quantify_uncertainty_pk_measure}
\end{figure}

In this Appendix, we quantify the systematic uncertainties in the power spectrum modeling discussed in Section~\ref{sec:modeling}. We first compared the SZ fluctuation power spectrum measured after performing Monte Carlo simulations to that predicted using the framework discussed in Section~\ref{sec:Relation_3D_2D_power_spectra}, in order to test the projection assumptions (Equation~\ref{eq:projection_only_modeling}). We then compared the power spectrum obtained from Monte Carlo realizations, including the instrument response function and data weighting, to that predicted from the modeling methodology discussed in Section~\ref{sec:modeling} in order to test the overall uncertainties associated with the modeling of the pressure fluctuation power spectrum.

Figure~\ref{fig:quantify_uncertainty_pk_measure} presents the comparison between the Monte Carlo and the model for projection only (top) and accounting for instrumental and weighting effects (bottom). The comparison is performed assuming a morphologically disturbed \cite{Arnaud2010} pressure profile, but we also show the impact of changing this model to a cool-core profile. The fluctuations were set to $\sigma_{\mathcal{P}} = 0.5$, $\alpha = -11/3$, $L_{\rm dis} = 1$ kpc, and $L_{\rm inj}$ was varied to change the power spectrum shape since this is key for the comparison. As we can see, the model predictions are accurate within $\sim10$\% at scales above or near the peak, $k \gtrsim 1/L_{\rm inj}$. At large scales, we observe larger deviations for $k << 1/L_{\rm inj}$, as expected (see Section~\ref{sec:Relation_3D_2D_power_spectra}). These deviations are larger for more peaked profiles, such as the cool-core profile in the test shown here. However, they appear in a regime where the power spectrum drops significantly with respect to the peak value and where it cannot be measured in practice.

We conclude that systematic uncertainties in the power spectrum modeling are negligible compared to other uncertainties in the present work. While we only show the case of a reference spectrum and profile, we have checked that these results are robust against the choice of the model and analysis parameters. We also note that the methodology developed here significantly improves over currently used derivations of the power spectrum, which essentially rely on the delta variance framework from \cite{Arevalo2012}. For more details, we point the reader to the extensive discussions in \cite{Romero2024b}, who show that power spectrum modeling uncertainties can reach a factor of a few, depending on the input power spectrum.

\section{Systematic uncertainties associated with the radial model}\label{app:radial_model_implication}
\begin{figure}
        \centering
        \includegraphics[width=0.49\textwidth]{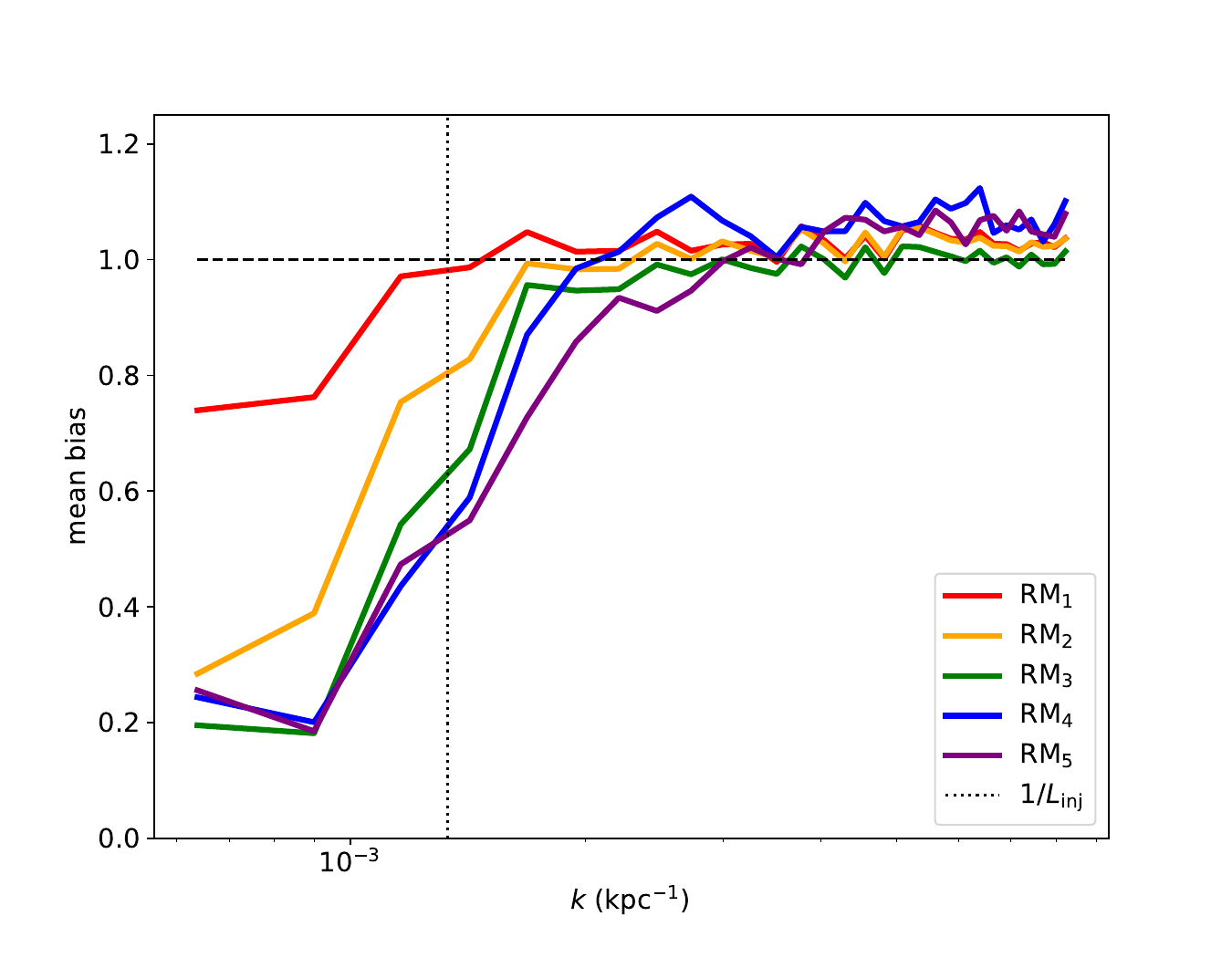}
        \includegraphics[width=0.49\textwidth]{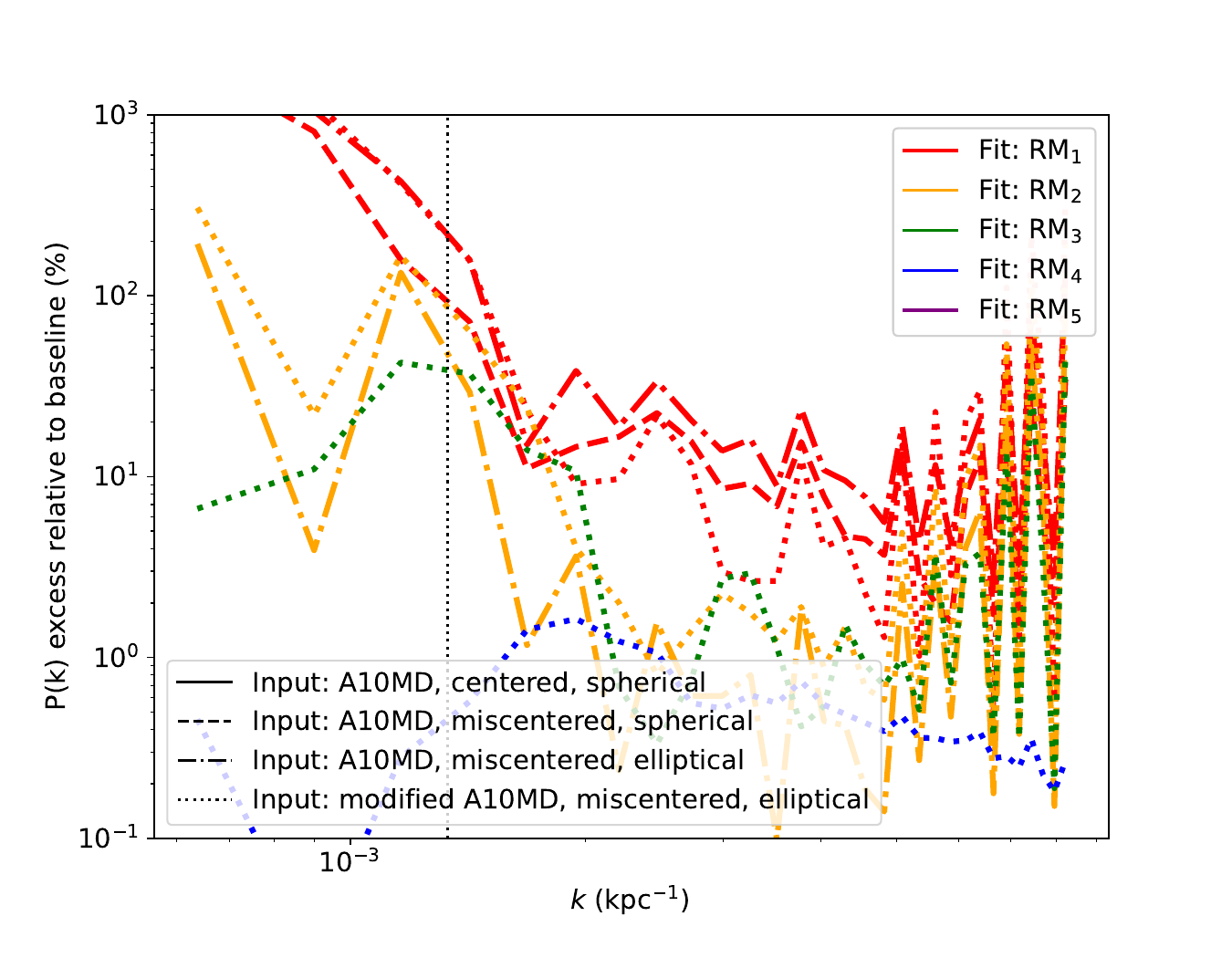}
        \caption{Estimate of the bias induced by the radial model as a function of wavenumber.
        {\bf Top}: bias induced by the removal of fluctuations implied by the fit of the radial model (Equation~\ref{eq:bias_overfit}).
        {\bf Bottom}: excess fluctuations induced by the mismodeling, relative to a reference pressure fluctuation power spectrum ($\sigma_{\mathcal{P}} = 0.5$, $L_{\rm inj} = 750$ kpc, $\alpha = -11/3$ and $L_{\rm dis} = 1$ kpc). 
        The different colors represent the different models, according to Section~\ref{sec:pressure_profile}. In the right panel, the different line styles refer to the different input radial model.
        In both panels, the vertical dashed line gives $k = 1/L_{\rm inj}$.
        }
\label{fig:test_radial_model_fit_syst}
\end{figure}

The systematic effects associated with the radial model are twofold. Firstly, the chosen radial model may not  adequately represent the cluster, such that differences will induce spurious SZ fluctuations. Secondly, fitting the radial model to the data is likely to absorb part of the fluctuations. In this Appendix, we attempt to estimate the magnitude of these systematic effects. 

\paragraph{Fluctuation removal from the radial model fit}

The first test consists of simulating SZ images, fitting the radial component according to the different models used in this paper, and comparing the resulting fluctuation power spectra to the one expected. Since any change in the radial model will also affect the conversion from the SZ 2D to the pressure 3D power spectrum, the SZ fluctuation power spectrum has to be normalized by $C_{3D \rightarrow 2D}^{({\rm eff})}$. Instrumental and masking effects are neglected because they affect all the simulated data in the same way. The induced bias is thus computed as
\begin{equation}
{\rm bias} = \frac{\left(\mathcal{P}_{\omega \delta y / \bar{y}} / C_{3D \rightarrow 2D}^{({\rm eff})}\right)_{\rm fitted \ RM}}{\left(\mathcal{P}_{\omega \delta y / \bar{y}} / C_{3D \rightarrow 2D}^{({\rm eff})}\right)_{\rm true \ RM}}.
\label{eq:bias_overfit}
\end{equation}

In the top panel of Figure~\ref{fig:test_radial_model_fit_syst}, we show the mean bias induced by radial model fitting to the data, as a function of $k$, obtained by averaging 100 mock realizations. The model used to generate the noiseless mocks was set to a spherical, morphologically disturbed \citep{Arnaud2010} pressure profile ($M_{500} = 2.5 \times 10^{15}$ M$_{\odot}$), and the pressure fluctuations were defined as lognormal with $\sigma_{\mathcal{P}} = 0.5$, $L_{\rm inj} = 750$ kpc, $\alpha = -11/3$ and $L_{\rm dis} = 1$ kpc. The fitted  radial models are those discussed in Section~\ref{sec:pressure_profile}. Note that in the context of this test, the models should perfectly fit the mocks in the absence of fluctuations. We can see that increasing the model complexity implies an increasing removal of fluctuations. Most of the filtering appears on large scales since the models are relatively smooth. Up to about 50\% filtering is observed near the power spectrum peak and up to 80\% on larger scales, for the most complex models. The bias converges to $\sim 1$ on small scales.

\paragraph{Incorrect radial model-induced bias}

Quantification of the impact of an incorrect radial model choice on the results is more difficult, since the true model of MACS~J0717.5+3745 is unknown. Moreover, the excess fluctuation bias should be quantified relative to a given fluctuation model. We thus proceeded as follows. We simulated SZ data given a radial model, without fluctuations, fit the image with another (different) radial model, and computed the SZ fluctuation power spectrum of the residuals. We then compare the induced spurious surface brightness fluctuations to a reference pressure fluctuation model given by $\sigma_{\mathcal{P}} = 0.5$, $L_{\rm inj} = 750$ kpc, $\alpha = -11/3$ and $L_{\rm dis} = 1$ kpc. The input radial models were defined in terms of  increasing complexity: 1) a centered, spherical, morphologically disturbed \cite{Arnaud2010} profile ; 2) as input model 1, but miscentered by 20 arcsec along R.A. and 20 arcsec along Dec. (this compares well with the difference between the X-ray and SZ peak for MACS~J0717.5+3745); 3) as input model 2, but with an elliptical profile with $q_{\rm int, maj}=0.7$; and 4) as input model 3, but changing the slopes of the pressure profile with $a \rightarrow a+0.1$, $b \rightarrow b-0.7$, $c \rightarrow c-0.2$, which typically corresponds to the difference one may have between different clusters \citep[e.g.,][]{Arnaud2010,PlanckV2013}.

In Figure~\ref{fig:test_radial_model_fit_syst}, bottom panel, we can observe that the excess power generally increases with input model complexity, and decreases with fit model complexity. This is expected since more complex models are better able to account for a mismatch with the data. The bias is also more prominent on large scales, at least given the tested models. As expected, no bias is observed when the model is able to fit the data perfectly. For instance, this is the case for the first input model, regardless of the fit model. 

According to this test, RM$_1$ may lead to an excess in the power spectrum larger than 10 times the reference power spectrum on large scales, because of miscentering. RM$_2$ may lead to a bias which is of the same order of magnitude as the reference spectrum, essentially because it does not account for ellipticity.  RM$_3$ may lead to up to 40\% bias due to the mismatch of the shape of the pressure profile (given the different shapes for the input and the fit used here). RM$_4$ is able to mitigate the mismatch in the shape by fitting $P_0$ and $r_p$ so that the bias reaches 1\% at most. Finally, RM$_5$ is able to perfectly match all the input pressure profiles so that it does not lead to any bias. Although these tests give an insight into the level of bias induced by the choice of the radial model, we stress that they strongly depend on the choice of the input models, and should only be taken as a qualitative estimate. 

Nevertheless, these two tests indicate that systematic effects in the choice of the radial model are likely to alter the measured power spectrum on scales comparable to or larger to the peak, and may thus alter the recovery of the peak power spectrum (e.g., $L_{\rm inj}$, $A(k_{\rm peak})$). Given the results obtained here, RM$_3$ appears as a reasonable baseline choice since the induced excess power expected compares well to the filtering, so that in the end the associated systematic effects may be minimal.

\section{Systematic uncertainties associated the kinetic Sunyaev-Zel'dovich signal}\label{app:kSZ_systematics}
\begin{figure*}
        \centering
        \includegraphics[width=\textwidth]{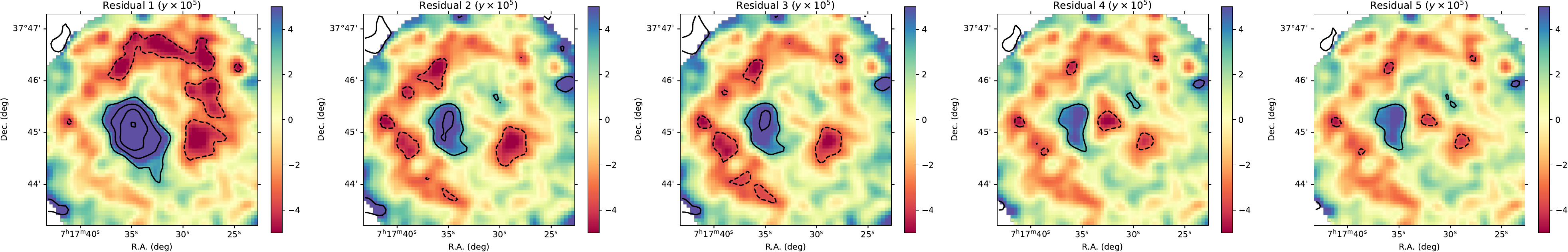}
        \caption{Same as Figure~\ref{fig:radial_model_residuals} with kSZ correction included.}
\label{fig:radial_model_residuals_ksz}
\end{figure*}

\begin{figure}
        \centering
        \includegraphics[width=0.49\textwidth]{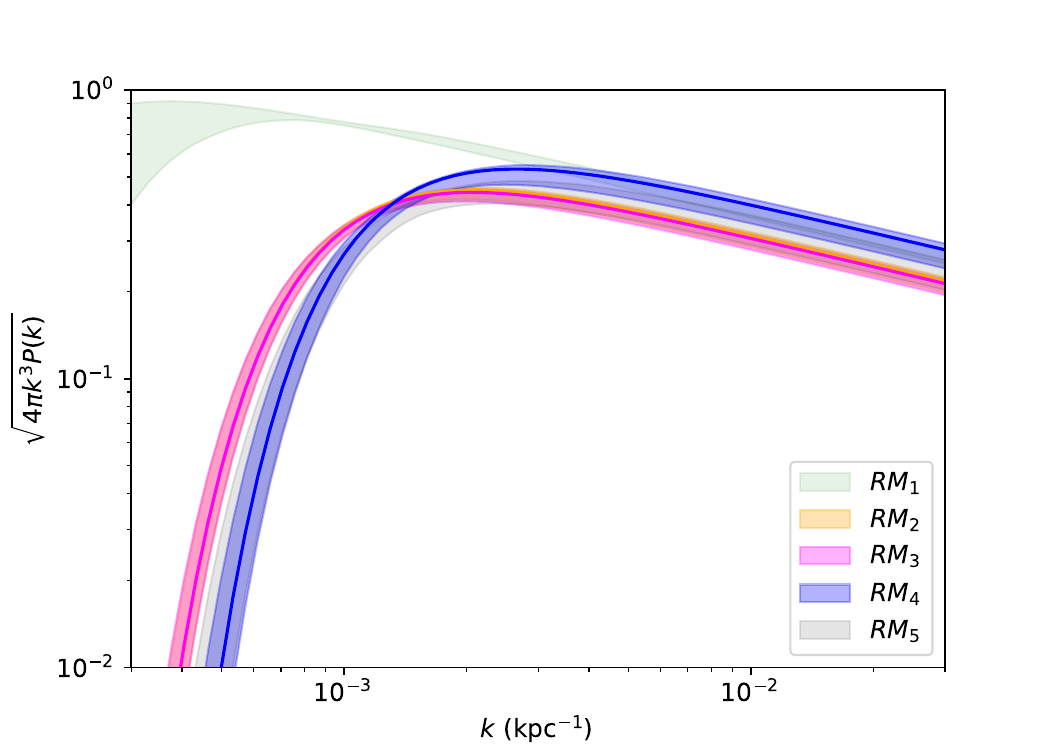}
        \caption{Same as Figure~\ref{fig:pk} right panel with kSZ correction included.
        }
\label{fig:pkksz}
\end{figure}

MACS~J0717.5+3745 is known to host significant kSZ signal. Following the recommendation from the public release of NIKA SZ data\footnote{\url{https://lpsc.in2p3.fr/NIKA2LPSZ/nika2sz.release.php}}, we used the kSZ corrected map to test the stability of the results. Figure~\ref{fig:radial_model_residuals_ksz} displays the residuals between the data and the considered models, as in Figure~\ref{fig:radial_model_residuals}, but accounting for the kSZ correction prior the fit. Similarly, Figure~\ref{fig:pkksz} shows the 3D pressure fluctuation power spectra accounting for the kSZ correction. The residual and recovered spectra are comparable to those of the baseline results, although model RM$_5$ is not able to fit the map as well as in the earlier version, and its resulting power spectra aligns better with RM$_{2}$, RM$_{3}$ and RM$_{4}$. The results of the power spectrum constraints are listed in Table~\ref{tab:inference_fluctuation_results}.

\end{appendix}

\end{document}